\documentclass[aps,pra,twocolumn,showpacs,floatfix]{revtex4-1}
\usepackage{graphicx,amsfonts,amssymb,amsmath,hyperref}
\usepackage{textcomp}
\usepackage{esint}

\newif\ifhyper
\hypertrue
\ifhyper
\hypersetup{
  citecolor = {green},
  colorlinks = {true}, 
  urlcolor = {blue} 
}  

\begin{document}

\graphicspath{{./figures_submit/}}


\def\lvdw{l_{\rm vdW}}
\def\Tkt{T_{\rm BKT}}
\def\gbh{g_{\rm BH}} 

\def\llbrace{\left\lbrace}
\def\rrbrace{\right\rbrace}

\newcommand{\Tr}{{\rm Tr}} 
\newcommand{\tr}{{\rm tr}} 
\newcommand{\sgn}{{\rm sgn}} 
\newcommand{\mean}[1]{\langle #1 \rangle}
\newcommand{\const}{{\rm const}} 
\newcommand{\cc}{{\rm c.c.}} 
\newcommand{\hc}{{\rm h.c.}} 

\def\eps{\epsilon}
\def\gam{\gamma} 
\def\phibf{\boldsymbol{\phi}}
\def\varphibf{\boldsymbol{\varphi}}
\def\psibf{\boldsymbol{\psi}}
\def\lamb{\lambda}
\def\Lamb{\Lambda}

\def\half{\frac{1}{2}}
\def\dhalf{\dfrac{1}{2}}
\def\third{\frac{1}{3}} 
\def\quarter{\frac{1}{4}}

\def\q{{\bf q}}
\def\r{{\bf r}}

\def\nablabf{\boldsymbol{\nabla}}
\def\rhobf{\boldsymbol{\rho}}
\def\sigmabf{\boldsymbol{\sigma}} 

\def\w{\omega}
\def\wn{\omega_n}
\def\wnu{\omega_\nu}
\def\dl{{\partial_l}}  
\def\dt{\partial_t} 
\def\tdt{\tilde\partial_t}
\def\dk{\partial_k}
\def\tdk{\tilde\partial_k}
\def\dx{\partial_x}
\def\dy{\partial_y} 
\def\dtau{{\partial_\tau}} 

\def\intq{\int_{\bf q}} 
\def\intr{\int d^dr}  
\def\dinttau{\displaystyle \int_0^\beta d\tau}
\def\dinttaup{\displaystyle \int_0^\beta d\tau'}
\def\inttau{\int_0^\beta d\tau}
\def\inttaup{\int_0^\beta d\tau'}
\def\intx{\int d^{d+1}x} 
\def\intw{\int_{-\infty}^\infty \frac{d\w}{2\pi}}

\def\dt{\partial_t}
\def\calD{{\cal D}}
\def\calF{{\cal F}} 
\def\calG{{\cal G}}
\def\calH{{\cal H}}
\def\calJ{{\cal J}}
\def\calL{{\cal L}} 
\def\calN{{\cal N}}
\def\calO{{\cal O}}
\def\calP{{\cal P}}  
\def\calR{{\cal R}} 
\def\calS{{\cal S}}

\def\tT{{\tilde T}}
\def\tmu{{\tilde\mu}}
\def\trho{{\tilde\rho}}


\title{Universal thermodynamics of a two-dimensional Bose gas} 

\author{A. Ran\c{c}on and  N. Dupuis}
\affiliation{Laboratoire de Physique Th\'eorique de la Mati\`ere Condens\'ee, 
CNRS UMR 7600, \\ Universit\'e Pierre et Marie Curie, 4 Place Jussieu, 
75252 Paris Cedex 05,  France}

\date{May 9, 2012}

\begin{abstract} 
Using renormalization-group arguments we show that the low-temperature thermodynamics of a three- or two-dimensional dilute Bose gas is fully determined by a universal scaling function $\calF_d(\mu/k_BT,\tilde g(T))$ once the mass $m$ and the $s$-wave scattering length $a_d$ of the bosons are known ($d$ is the space dimension). Here $\mu$ and $T$ denote the chemical potential and temperature of the gas, and the temperature-dependent dimensionless interaction constant $\tilde g(T)$ is a function of $ma_d^2k_BT/\hbar^2$. We compute the scaling function $\calF_2$ using a nonperturbative renormalization-group approach and find that both the $\mu/k_BT$ and $\tilde g(T)$ dependencies are in very good agreement with recent experimental data obtained for a quasi-two-dimensional Bose gas with or without optical lattice. We also show that the nonperturbative renormalization-group estimate of the Berezinskii-Kosterlitz-Thouless transition temperature compares well with the result obtained from a quantum Monte Carlo 
simulation of an effective classical field theory.   
\end{abstract}
\pacs{05.30.Rt,05.30.Jp,67.85.-d,03.75.Hh}
\maketitle

\section{Introduction}

The physical properties of a system at thermal equilibrium are determined by an equation of state. For a fluid of particles in the grand canonical ensemble, the equation of state relates a thermodynamic quantity such as pressure, density or entropy to temperature and chemical potential. It can take a complicated expression when the particles interact {\it via} a two-body potential $V(\r_1-\r_2)$ which has no simple expression as is usually the case in real systems. Quite remarkably however, the low-temperature equation of state of a $d$-dimensional dilute gas is universal, in the sense that it depends only on a small number of parameters, such as the mass $m$ of the particles and the $s$-wave scattering length $a_d$, and is otherwise insensitive to the details of the two-body potential $V(\r_1-\r_2)$. A well-know example of universality is given by a three-dimensional dilute Bose gas at zero temperature, the pressure of which is given by the mean-field result $P(\mu)=\mu^2m/(8\pi a_3\hbar^2)$ to leading 
order in the small parameter $ma_3^2\mu/\hbar^2$. The first quantum correction, known as the Lee-Huang-Yang correction, is also entirely determined by $m$ and $a_3$ (besides the chemical potential $\mu$)~\cite{Lee57a,*Lee57b}. (In the following we set $k_B=\hbar=1$.)

The universality property of the equation of state of a dilute Bose gas can be understood from the point of view of the theory of phase transitions~\cite{Sachdev_book,Fisher89,Zhou10,Hazzard11}. By varying the chemical potential from negative to positive values at zero temperature, one induces a quantum phase transition between a state with vanishing pressure and no particles (vacuum) and a superfluid state with a nonzero pressure. This identifies the point $\mu=T=0$ as a quantum critical point (QCP). Above two dimensions (the upper critical dimension of the $T=0$ quantum phase transition), {\it i.e.} for $d\geq 2$, the boson-boson interaction is irrelevant and the critical behavior at the transition is mean-field like with a correlation-length exponent $\nu=1/2$ and a dynamical exponent $z=2$. However, the boson-boson interaction cannot be completely ignored and enters the equation of state~\cite{note2}. In the critical regime near the QCP, defined by $ml^2\mu\ll1$ and $ml^2T\ll 1$ with $l$ the ``natural" 
low-energy length scale~\cite{Braaten06,note15}, the pressure takes the form 
\begin{equation} 
P(\mu,T) = \left(\frac{m}{2\pi}\right)^{d/2} T^{d/2+1} \calF_d\left(\frac{\mu}{T},\tilde g(T)\right) ,
\label{pressure0}
\end{equation} 
where $\calF_d$ is a universal scaling function characteristic of the $d$-dimensional dilute Bose gas universality class. The temperature-dependent dimensionless interaction constant $\tilde g(T)$ is a known function of $ma_d^2T$, so that $P(\mu,T)$ can also be written in terms of a universal function of $\mu/T$ and $ma_d^2T$. In two dimensions and in the weak-interaction limit, $\tilde g(T)\equiv \tilde g$ is approximately temperature independent and the universal scaling function $\calF_2(\mu/T,\tilde g)$ depends on $\mu/T$ with the interaction strength $\tilde g$ as a parameter; the equation of state then exhibits an approximate scale invariance (with no characteristic energy scales other than $\mu$ and $T$)~\cite{Prokofev01,Prokofev02}. Equation~(\ref{pressure0}) also holds in a one-dimensional Bose gas ({\it i.e.} below the upper critical dimension of the $T=0$ vacuum-superfluid transition) but with the universal function $\calF_1$ depending only on $\mu/T$.

While Eq.~(\ref{pressure0}) follows from general renormalization-group (RG) arguments (see Sec.~\ref{sec_univ}), the theoretical determination of the universal scaling function $\calF_d(x,y)$ requires an explicit computation of the pressure $P(\mu,T)$. A perturbative calculation order by order in $\tilde g(T)$ is possible only for $d>2$ (it nevertheless breaks down in the critical regime of the thermal phase transition between the normal and the superfluid phase, which is controlled by the Wilson-Fisher fixed point of the classical O(2) model). In two dimensions, perturbative theory is plagued with infrared divergences at finite temperatures, thus making the determination of $\calF_2$ difficult, in particular in the quantum critical regime $|\mu|\ll T$. The Berezinskii-Kosterlitz-Thouless (BKT) transition~\cite{Berezinskii70,*Berezinskii71,*Kosterlitz73,*Kosterlitz74} and the low-temperature phase with quasi-long-range order are also beyond a mere perturbative treatment~\cite{Fisher88}.  

The advantage of the point of view based on phase transitions is two-fold. Firstly it gives a straightforward explanation of universality in a dilute Bose gas. Secondly it shows that the universal equation of state~(\ref{pressure0}) holds not only for a dilute Bose gas but for any system near a quantum phase transition belonging to the same universality class. For instance, a Bose gas in an optical lattice near the vacuum-superfluid transition exhibits the same thermodynamics as a dilute Bose gas, provided that $m$ and $a_d$ are understood as the effective mass and scattering length of the bosons moving in the lattice. The thermodynamics of a Bose gas near the superfluid--Mott-insulator transition is also described by the equation of state~(\ref{pressure0}), since this quantum phase transition (when it is induced by a density change) belongs to the dilute Bose gas universality class~\cite{Sachdev_book,Fisher89}. In this manuscript we focus on the vacuum-superfluid transition in a two-dimensional Bose gas. 

On the experimental side, cold atomic gases provide us with highly controlled and tunable systems where universal thermodynamics can be experimentally demonstrated. Altough cold gases are inhomogeneous and of finite size due to the harmonic confining potential, using a local-density approximation it is possible to deduce the equation of state $P(\mu,T)$ of the infinite homogeneous gas (with uniform density)~\cite{Cheng07,Ho09}. A number of experiments on weakly-interacting two-dimensional Bose gases have been reported~\cite{Hadzibabic06,Clade09,Rath10,Hung11,Yefsah11}, and the scale invariance of the equation of state $P(\mu,T)$ has been observed~\cite{Rath10,Hung11,Yefsah11}. More recently, the equation of state of a Bose gas in an optical lattice has been measured near the vacuum-superfluid transition in a regime where the interaction constant is not weak~\cite{Zhang12}. These experiments allow us to determine both the $\mu/T$ and $\tilde g(T)$ dependence of the universal scaling function $\calF_2$ in 
various limits and will be thoroughly discussed in the manuscript. 

The outline of the paper is as follows. In Sec.~\ref{sec_model}, we introduce and motivate the low-energy effective Hamiltonians which enable to derive the universal thermodynamics of three- and two-dimensional dilute Bose gases. Section~\ref{sec_univ} is devoted to a discussion of the thermodynamics of a dilute Bose gas using the language and concepts familiar from the theory of phase transitions. A detailed derivation of Eq.~(\ref{pressure0}) is given. In Sec.~\ref{sec_F}, we discuss the universal scaling function $\calF_2$ obtained from a nonperturbative renormalization-group (NPRG) approach. These theoretical results are compared with the experimental data of Refs.~\cite{Yefsah11,Hung11,Zhang12} in Sec.~\ref{sec_exp}. In particular, we make quantitative comparisons between the experimental data obtained with a Bose gas in an optical lattice~\cite{Zhang12} and theoretical results obtained in the framework of the Bose-Hubbard model. Finally, in Sec.~\ref{sec_bkt}, we discuss the NPRG prediction for the BKT 
transition temperature and compare it with the estimate deduced from a quantum Monte Carlo simulation of an effective classical field theory.

\section{Model Hamiltonians} 
\label{sec_model}

\subsection{Three-dimensional Bose gas} 

The interaction between ultracold atoms is governed by a potential $V(\r_1-\r_2)$ which is repulsive at short distances and determined by the van der Waals attraction $-C_6|\r_1-\r_2|^{-6}$ at long distances~\cite{Bloch08,Braaten06}. The latter defines the microscopic length scale $\lvdw\sim (mC_6)^{1/4}$ ($m$ denotes the atomic mass). For length scales larger than $\lvdw$ and energies smaller than $1/m\lvdw^2$, collisions between atoms occur only in the $s$-wave channel and the scattering amplitude is well approximated by 
\begin{equation}
f_{\rm 3D}(\q) = - \frac{a_3}{1+i|\q|a_3} , 
\label{scat3D}
\end{equation}
where the three-dimensional $s$-wave scattering length $a_3$ is typically of the order of $\lvdw$. In this low-energy regime, the ultracold gas can be described by the effective Hamiltonian 
\begin{align}
\hat H ={}& \intr \biggl\lbrace \hat\psi^\dagger(\r) \left(-\frac{\nablabf^2}{2m}-\mu \right)\hat\psi(\r) \nonumber \\ & + \frac{g}{2}\hat\psi^\dagger(\r)\hat\psi^\dagger(\r) \hat\psi(\r)\hat\psi(\r) \biggr\rbrace , 
\label{ham}
\end{align}
with an ultraviolet momentum cutoff $\Lambda\sim \lvdw^{-1}$ ($d=3$ for a three-dimensional gas). Here $\hat\psi^{(\dagger)}(\r)$ is a bosonic operator, $\mu$ the chemical potential and $g$ the ``microscopic'' interaction constant. The scattering amplitude obtained from~(\ref{ham}) takes the form~(\ref{scat3D}) with a scattering length
\begin{equation}
a_3 = \frac{mg}{4\pi+\frac{2}{\pi}mg\Lambda}  
\end{equation}
which is a function of $g$ and $\Lambda$. The low-energy effective description is valid only for momentum scales much smaller than $\Lamb$, which requires both temperature and density to be small enough: $T\ll \Lambda^2/2m$ and $D\ll \Lambda^3$. 

\subsection{Two-dimensional Bose gas} 
\label{subsec_Q2D} 

A quasi-two-dimensional gas can be created by subjecting a three-dimensional gas to a confining harmonic potential along one direction. The scattering amplitude then vanishes in the low-energy limit $\q\to 0$, 
\begin{equation}
f_{\rm 2D}(\q) = - \frac{2\pi}{\ln\left(\frac{|\q|a_2}{2}\right)+C-i\frac{\pi}{2}} 
\label{scat2D}
\end{equation}
($C$ is the Euler constant), as in a strictly two-dimensional system~\cite{Petrov00a,Lim08}. The effective two-dimensional $s$-wave scattering length $a_2$ is a function of the thickness $l_z$ of the gas in the confining direction, as well as the $s$-wave scattering length and microscopic interaction strength of the three-dimensional (unconfined) Bose gas. At sufficiently low temperatures, when $T$ is much smaller than the $\w_z=1/ml_z^2$, only the lowest level of the confining potential is populated and the gas behaves as a two-dimensional system. The quasi-two-dimensional gas can be described by the effective Hamiltonian~(\ref{ham}) with $d=2$ and a ``microscopic'' interaction constant~\cite{Petrov00a,Lim08}
\begin{equation}
g = \sqrt{8\pi} \frac{a_3}{ml_z} ,
\end{equation}
which reproduces the scattering amplitude~(\ref{scat2D}) with the scattering length
\begin{equation}
a_2 = \frac{2}{\Lambda} \exp\left( -\frac{2\pi}{mg}-C \right) . 
\label{a2}
\end{equation}
Here $\Lambda\sim l_z^{-1}$ is an ultraviolet momentum cutoff below which the two-dimensional description holds. In addition to the condition $T\ll\w_z\sim \Lambda^2/2m$, we must require the density to satisfy $D\ll\Lambda^2$. The typical energy per particle $gD$ is then much smaller than $\w_z$ as it should be for the two-dimensional description to be justified. Note that in all experiments realized so far, the dimensionless interaction constant $\tilde g=2mg$ is small.

\subsection{Bose gas in an optical lattice}

Bosons in an optical lattice are described by the Bose-Hubbard model~\cite{Jaksch98,Fisher89}, 
\begin{equation}
\hat H = - t \sum_{\mean{\r,\r'}} \left( \hat\psi^\dagger_\r \hat\psi_{\r'} + \hc \right) + \sum_\r \left[ -\mu \hat n_\r + \frac{U}{2} \hat n_\r (\hat n_\r-1) \right] , 
\label{BH}
\end{equation}
where $t$ is the hopping amplitude between nearest-neighbor sites $\mean{\r,\r'}$, and $U$ the onsite interaction. $\hat\psi_\r^{(\dagger)}$ is an annihilation (creation) operator for a boson at site $\r$ of the lattice and $\hat n_\r=\hat\psi^\dagger_\r \hat\psi_\r$. An effective single-band description is valid only if the optical potential is strong enough and at sufficiently low temperatures. For a $d$-dimensional hypercubic lattice, the dispersion of the free bosons is given by the Fourier transform $t_\q = -2t\sum_i \cos(q_il)$ of the intersite hopping matrix ($l$ denotes the lattice spacing). It is convenient to use a shifted dispersion law 
\begin{equation}
\eps_\q = 2t d -2t \sum_{i=1}^d \cos(q_il)  
\label{epsq} 
\end{equation}
which vanishes for $\q=0$ and behaves as $\eps_\q\simeq tl^2\q^2$ for $|\q|\ll l^{-1}$. 

If the density $D$ is low enough ($Dl^d\ll 1$), the ground state is always a superfluid and we do not have to worry about the physics of the Mott transition~\cite{Fisher89}. Furthermore, at low temperatures $T\ll t$, the lattice does not matter and one can take the continuum limit where the Hamiltonian takes the form~(\ref{ham}) with an effective mass $m=1/2tl^2$, an interaction constant $g=Ul^d$, and a chemical potential $\mu+2dt$. The ultraviolet momentum cutoff $\Lambda$ is of the order of the inverse lattice spacing $l^{-1}$; the conditions $T\ll t$ and $Dl^d\ll 1$ then become $T\ll\Lambda^2/2m$ and $D\ll\Lambda^d$.

Thus, in the low-energy limit, a Bose gas in an optical lattice behaves similarly to a homogeneous Bose gas with an effective mass $m$ and an effective interaction constant $g$. To ensure that the effective continuum model reproduces the same low-energy physics as the lattice model, we must choose the cutoff $\Lambda$ so that it yields the same scattering length. In the two-dimensional case, we require Eq.~(\ref{a2}) to reproduce the scattering length of the two-dimensional Bose-Hubbard model~\cite{note9},  
\begin{equation}
a_2 = \frac{l}{2\sqrt{2}} \exp\left(-\frac{4\pi t}{U} - C\right) ,
\label{a2lat}  
\end{equation}
which gives $\Lambda=4\sqrt{2}/l$.                                                                                                                                                                                                                                                                                                                           

\section{Universal thermodynamics} 
\label{sec_univ}

In this section, we discuss the thermodynamics of a $d$-dimensional dilute Bose gas from the point of view of phase transitions, starting from the Hamiltonian~(\ref{ham}). This description provides us with a natural explanation of universality as well as a simple derivation of Eq.~(\ref{pressure0}). Altough we will mainly focus on two-dimensional systems in the following sections, for generality we consider an arbitrary dimension $d\geq 2$. 

\subsection{Vacuum-superfluid transition}

Let us first consider the vacuum-superfluid quantum phase transition induced by a change of chemical potential at zero temperature. For $d$ larger than the upper critical dimension $d_c^+=2$, boson-boson interactions are irrelevant (in the RG sense) and the critical behavior is described by non-interacting bosons. At the QCP $\mu=0$, the ground state is the vacuum, and the single-particle Green function is given by
\begin{equation}
G(\q,i\w) = \left( i\w - \frac{\q^2}{2m} \right)^{-1} , 
\label{propa}
\end{equation}
with $\w$ a (bosonic) Matsubara frequency. This result is exact and holds for any value of the (bare) interaction constant $g$~\cite{note1}. We deduce the dynamic exponent $z=2$ while the anomalous dimension $\eta$ vanishes. Similarly, for $\mu\leq 0$, we find $G(\q,i\w)^{-1} = i\w+\mu -\q^2/2m$ and the critical exponent associated with the correlation length $\xi=|2m\mu|^{-1/2}$ takes the value $\nu=1/2$. The value of the renormalized interaction $g_R$ at the QCP is given by the $T$ matrix in vacuum (using again the fact that the ground state is the vacuum). In the low-energy limit, it takes the value $4\pi a_3/m$ in three dimensions, but vanishes logarithmically in two dimensions (see Eq.~(\ref{gtilde}) below). 

The same analysis holds for bosons moving in a lattice. In the vacuum, the single-particle Green function is given by 
\begin{equation}
G(\q,i\w) = (i\w +\mu+2dt- \eps_\q)^{-1} , 
\end{equation}
where $\eps_\q$ is the dispersion of the free bosons [Eq.~(\ref{epsq})]. The $T=0$ QCP between the vacuum and the superfluid phase is now located at $\mu_c=-2dt$ and the elementary excitations have an effective mass $m=1/2tl^2$. As in the continuum model, the renormalized value $g_R$ of the interaction ({\it i.e.} the $T$ matrix in vacuum) can be expressed in terms in of the scattering length $a_d$ of the bosons moving in the lattice~\cite{Rancon11b}.

\subsection{RG approach} 

The preceding results can be formulated in the language of the RG. In the Wilson formulation, a RG transformation consists in integrating out ``fast'' modes with momenta between $\Lambda$ and $\Lambda/s$ ($s>1$), and rescaling fields, momenta and frequencies in order to restore the original value of the cutoff $\Lambda$. This yields an effective Hamiltonian for the ``slow'' modes with a renormalized interaction constant $g(s)$~\cite{note16}. At the QCP $\mu=T=0$, there is no renormalization of the quadratic part of the Hamiltonian, in agreement with the fact that Eq.~(\ref{propa}) is exact. The dimensionless interaction constant $\tilde g(s)=2m \Lambda^{d-2}g(s)$ satisfies the RG equation
\begin{equation}
s \frac{d\tilde g(s)}{ds} = (2-d)\tilde g(s) - \frac{K_d}{2} \tilde g(s)^2 ,
\label{rgeq1} 
\end{equation}
where $K_d=[2^{d-1}\pi^{d/2}\Gamma(d/2)]^{-1}$. Above the upper critical dimension $d_c^+=2$, $\tilde g(s)$ vanishes for $s\to\infty$ and the only fixed point of Eq.~(\ref{rgeq1}) is the Gaussian fixed point $\tilde g=0$, which therefore governs the quantum phase transition between the vacuum and the superfluid phase. From~(\ref{rgeq1}), we obtain 
\begin{equation}
\tilde g(s) = \llbrace 
\begin{array}{lcc} 
\dfrac{8\pi\Lambda a_3}{s} & \mbox{if} & d=3 , \\
- \dfrac{4\pi}{\ln\left(\frac{\Lambda a_2}{2s}\right)+C} & \mbox{if} & d=2 , 
\end{array}
\right.
\label{gtilde}
\end{equation}
where the result for $d=3$ holds for $s\gg 1$. 

\begin{figure}
\centerline{\includegraphics[width=6.5cm]{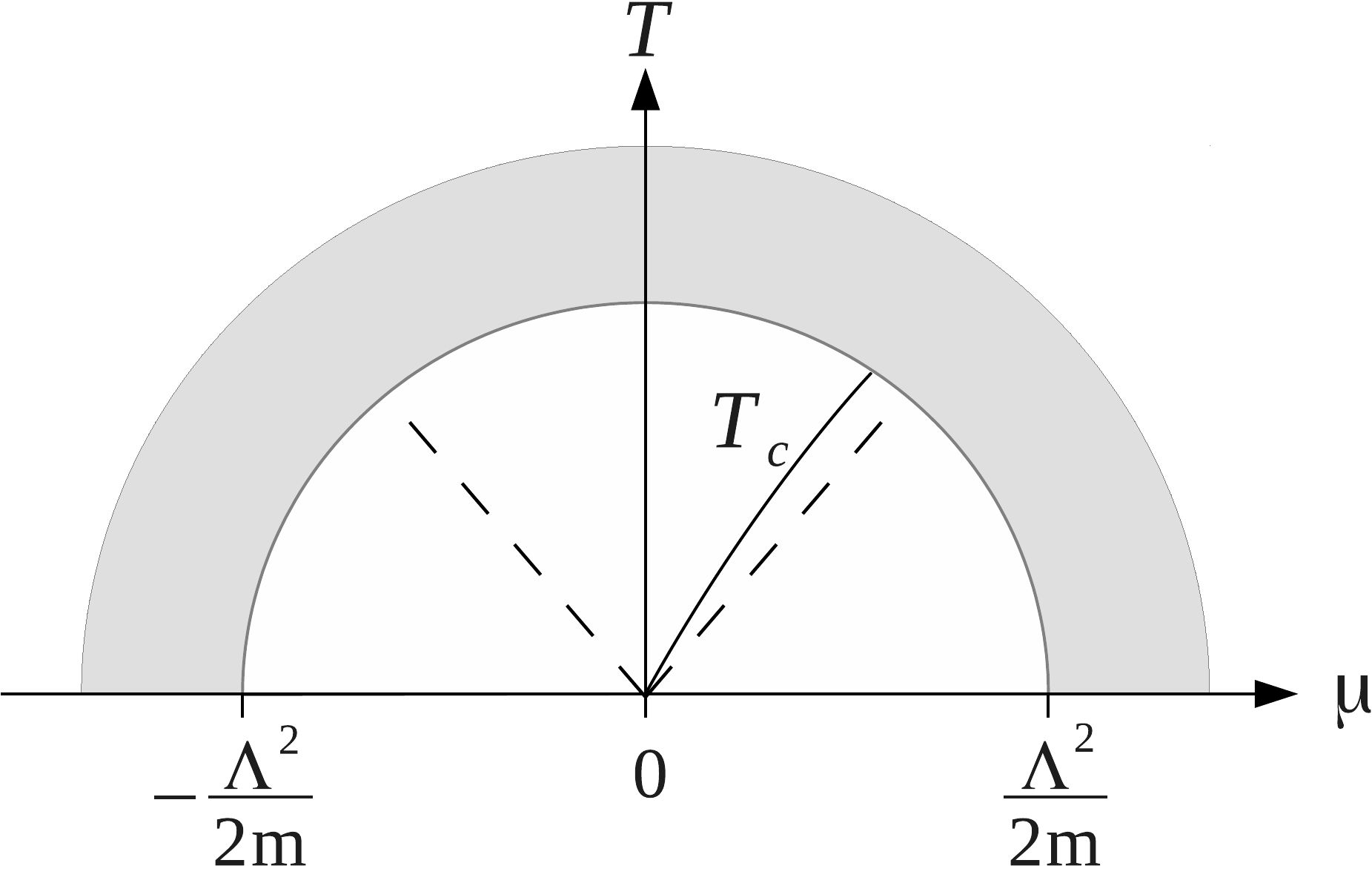}}
\caption{(Color online) Phase diagram of the dilute Bose gas ($d\geq 2$). The dashed lines are defined by $|\mu|\sim T$ and the solid one corresponds to the superfluid transition (of BKT type when $d=2$). The shaded area corresponds to the high-energy region $|\mu|,T\gtrsim \Lambda^2/2m$ where the thermodynamics is not universal. The value of the ultraviolet momentum cutoff $\Lamb$ is discussed in Sec.~\ref{sec_model} for a three- and a quasi-two-dimensional Bose gas.}
\label{fig_phase_dia} 
\end{figure}

There are two relevant perturbations about the Gaussian fixed point $\mu=T=\tilde g=0$: the chemical potential and the temperature. In a RG transformation, they transform as 
\begin{equation}
\tilde T(s) = s^z \tilde T , \qquad
\tilde\mu(s) = s^{1/\nu} \tilde\mu ,
\label{rgeq2}
\end{equation}
near the QCP ({\it i.e.} when $|\tmu(s)|\lesssim 1$ and $\tilde T(s)\lesssim 1$). We have introduced the dimensionless variables~\cite{note3} 
\begin{equation}
\tilde T = \frac{2mT}{\Lambda^z}, \qquad \tilde\mu = \frac{2m\mu}{\Lambda^{1/\nu}} .
\end{equation}
Note that in the low-temperature regime where this analysis based on the effective Hamiltonian~(\ref{ham}) is valid, $|\tmu|\ll 1$ and $\tilde T\ll 1$ (see Sec.~\ref{sec_model}). When $\mu$ and $T$ are nonzero, the RG equation for $\tilde g(s)$ is well approximated by~(\ref{rgeq1}) or (\ref{gtilde}) as long as $|\tmu(s)|\lesssim 1$ and $\tilde T(s)\lesssim 1$. We can obtain a rough sketch of the phase diagram by noting that the low-energy behavior of the system depends on which of the conditions $|\tmu(s)|\sim 1$ and $\tilde T(s)\sim 1$ is reached first. This yields two crossover lines defined by $|\tmu|\sim \tilde T$, {\it i.e.} $|\mu|\sim T$ using $z=1/\nu=2$, in agreement with the generic phase diagram of a system near a quantum critical point (see Fig.~\ref{fig_phase_dia})~\cite{Sachdev_book}. For $\mu<0$ and $|\mu|\gg T$, the system behaves as a dilute classical gas and we expect a classical Boltzmann description to apply (see Sec.~\ref{sec_F}). The condition $|\mu|\ll T$ defines the quantum critical 
regime where the physics is controlled by the QCP $\mu=T=0$ and its thermal excitations~\cite{Sachdev_book}.

\subsection{Universal thermodynamics}

Let us now consider the dimensionless pressure~\cite{note3}
\begin{equation}
\tilde P(\tmu,\tT,\tilde g) = \frac{2m}{\Lambda^{d+z}} P(\mu,T) ,
\end{equation}
expressed in terms of the dimensionless variables $\tilde T$, $\tilde\mu$ and $\tilde g$ (note that $\tilde P$ has no explicit dependence on the ultraviolet cutoff $\Lambda$). In a RG transformation, $\tilde P$ transforms as
\begin{equation} 
\tilde P(\tmu,\tT,\tilde g) = s^{-d-z} \tilde P(s^{1/\nu}\tmu,s^z\tilde T,\tilde g(s)) ,
\label{pressure}
\end{equation}
provided that $\tilde T(s)\ll 1$ and $|\tmu(s)|\ll 1$. Equation~(\ref{pressure}) holds for the full pressure since the vanishing of $P$ when $\mu\leq 0$ and $T=0$ implies that $P$ has no regular part at the transition. Only the two-body interaction constant $\tilde g$ is taken into account. Higher-order interactions (which are inevitably generated in the RG procedure), such as the three-body term, are not considered here since they are irrelevant and give rise to subleading contributions to the pressure~\cite{note4,note8}

Setting $s=\tilde T^{-1/z}$ in Eq.~(\ref{pressure}), we obtain 
\begin{equation}
\tilde P(\tmu,\tT,\tilde g) = \frac{\tT^{d/z+1}}{(4\pi)^{d/2}} \calF_d\left(\frac{\tmu}{\tT^{1/\nu z}},\tilde g(\tT)\right) ,
\label{pressure2}
\end{equation}
where we use the notation $\tilde g(\tilde T)$ for $\tilde g(s=\tT^{-1/z})$. Going back to dimensionful variables and setting $z=1/\nu=2$, we finally obtain Eq.~(\ref{pressure0}) where the energy-dependent interaction constant 
\begin{equation}
\tilde g(\eps) = \llbrace 
\begin{array}{lcc}
8\pi \sqrt{2ma_3^2\eps} & \mbox{if} & d=3 , \\
- \dfrac{4\pi}{\ln\left(\half\sqrt{2ma_2^2\eps}\right)+C} & \mbox{if} & d=2 ,
\end{array}
\right.
\label{gT}
\end{equation}
is obtained from~(\ref{gtilde}) with $s=\tilde\eps^{-1/2}$ and $\tilde\eps=2m\eps/\Lambda^2$. We stress that $\calF_d$ is a universal scaling function characteristic of the $d$-dimensional dilute Bose gas universality class (the factor $1/(4\pi)^{d/2}$ in~(\ref{pressure2}) is introduced for later convenience); it is independent of microscopic parameters such as the mass $m$ of the bosons or the scattering length $a_d$ which depend on the system considered.  

Using Eq.~(\ref{pressure0}), we can write any thermodynamic quantity in a scaling form. For instance, the density $D=\partial P/\partial\mu$ and the entropy per unit volume $s=\partial P/\partial T$ read
\begin{equation}
D(\mu,T) =  \left(\frac{mT}{2\pi}\right)^{d/2} \calF_d^{(1,0)}\left(\frac{\mu}{T},\tilde g(T)\right) ,
\end{equation}
and
\begin{multline}
s(\mu,T) = \left(\frac{mT}{2\pi}\right)^{d/2} \biggl[ \left(\frac{d}{2}+1\right) \calF_d\left(\frac{\mu}{T},\tilde g(T)\right) \\ - \frac{\mu}{T} \calF_d^{(1,0)}\left(\frac{\mu}{T},\tilde g(T)\right) + T \tilde g'(T) \calF_d^{(0,1)}\left(\frac{\mu}{T},\tilde g(T)\right) \biggr] ,
\label{sdef}
\end{multline}
where $\calF_d^{(1,0)}(x,y)\equiv\partial_x\calF_d(x,y)$, $\calF_d^{(0,1)}(x,y)\equiv\partial_y\calF_d(x,y)$, and $\tilde g'(T)=d\tilde g/dT$. Since $T\tilde g'(T)=\frac{d-2}{2}\tilde g(T)$ for $d>2$ and $T\tilde g'(T)=\tilde g(T)^2/8\pi$ for $d=2$, $s(\mu,T)$ is a function of $\mu/T$ and $\tilde g(T)$ (up to the factor $(mT)^{d/2}$). 

One often introduces the so-called phase-space pressure and phase-space density,
\begin{equation}
\begin{split}
\calP(\mu,T) &= P(\mu,T) \frac{\lamb_{\rm dB}^d}{T} = \calF_d\left(\frac{\mu}{T},\tilde g(T)\right) ,\\
\calD(\mu,T) &= D(\mu,T) \lamb_{\rm dB}^d = \calF_d^{(1,0)}\left(\frac{\mu}{T},\tilde g(T)\right) ,
\end{split}
\label{calPD}
\end{equation}
where $\lamb_{\rm dB}=\sqrt{2\pi/mT}$ is the thermal de Broglie wavelength. $\calP$ and $\calD$ provides a direct measure of the scaling function $\calF_d$ and its derivative $\calF_d^{(1,0)}$. One can also consider the entropy per particle $\calS=D^{-1}\partial P/\partial T$, 
\begin{equation}
\calS(\mu,T) = -\frac{\mu}{T} + \left(\frac{d}{2}+1\right) \frac{\calF_d}{\calF_d^{(1,0)}} + T\tilde g'(T) \frac{\calF_d^{(0,1)}}{\calF_d^{(1,0)}} ,
\label{calS} 
\end{equation}
where we use the shorthand notation $\calF_d\equiv\calF_d(\mu/T,\tilde g(T))$, etc. Note that $\calS(\mu,T)$ is also a universal function of $\mu/T$ and $\tilde g(T)$ (see the remark about $T\tilde g'(T)$ following Eq.~(\ref{sdef})). 

At zero temperature, the scaling function $\calF_d$ can be computed in perturbation theory. The one-loop correction to the mean-field result gives 
\begin{equation}
P(\mu,0) = \Theta(\mu) \frac{m\mu^2}{8\pi a_3} \left(1 - \frac{64}{15\pi} \sqrt{ma_3^2\mu} \right) 
\end{equation}
in three dimensions (the one-loop correction is known as the Lee-Huang-Yang correction~\cite{Lee57a,*Lee57b}), and
\begin{equation}
P(\mu,0) = - \Theta(\mu) \frac{m\mu^2}{4\pi} \left[ \ln\left(\half\sqrt{ma_2^2\mu}\right)+C+\quarter \right] 
\label{P2}
\end{equation}
in two dimensions, where $\Theta$ denotes the step function~\cite{Schick71,Popov72a,Popov_book_2}. These results can be cast in the scaling form 
\begin{equation}
P(\mu,T) =  \left(\frac{m}{2\pi}\right)^{d/2} \mu^{d/2+1} \calG_d\left(\frac{T}{\mu},\tilde g(\mu)\right) ,
\label{calG}
\end{equation}
which is equivalent to Eq.~(\ref{pressure0}) but more appropriate to the zero-temperature limit~\cite{note5}. 

At finite temperature, the determination of the scaling function $\calF_2$ (or $\calG_2$) is difficult in two dimensions, in particular in the quantum critical regime $|\mu|\ll T$. In the following section, we discuss the scaling function $\calF_2$ obtained from the NPRG approach.

\section{Scaling function $\calF$ of a two-dimensional Bose gas}
\label{sec_F}

The NPRG approach has recently been used to understand the physics of a Bose gas beyond the Bogoliubov approximation~\cite{Andersen99,Wetterich08,Dupuis07,*Dupuis09a,*Dupuis09b,Floerchinger08,*Floerchinger09b,Floerchinger09a,Sinner09,*Sinner10,*Eichler09}, but the computation of the scaling function $\calF_d$ has  not been carried out except for the Lee-Huang-correction in a zero-temperature three-dimensional Bose gas~\cite{Floerchinger08,*Floerchinger09b}. Here we discuss the NPRG results for the scaling function $\calF\equiv\calF_2$ which determines the thermodynamics of a two-dimensional Bose gas. We use both the standard version of the NPRG as well as its lattice version~\cite{Rancon11a,Rancon11b} to directly study the Bose-Hubbard model. The NPRG approach is briefly reviewed in Appendix~\ref{app_nprg}. Our results are based on the numerical solution of the NPRG equations as well as analytical results in some limits, in particular for $\mu=0$ (see Appendix~\ref{app_muzero}).  

\subsection{$\calF(x,y)$ vs $x$ ($y$ fixed)}
\label{subsec_Fx} 

\begin{figure}
\centerline{\includegraphics[width=7cm]{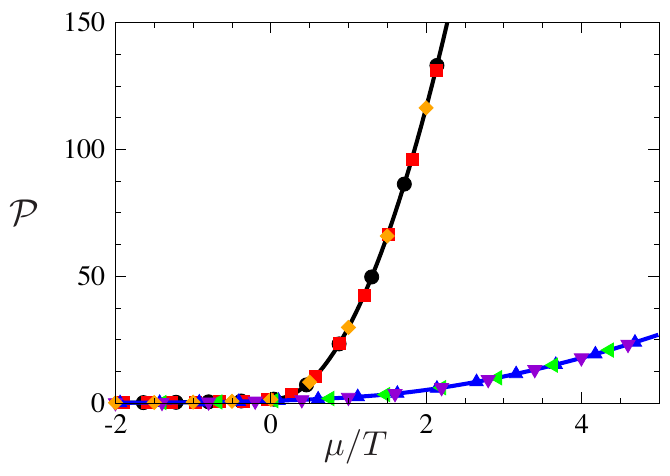}}
\caption{(Color online) Phase-space pressure $\calP(\mu,T)$ vs $\mu/T$ at fixed $\tilde g(T)$. The upper symbols are obtained for $\tilde g(T)=0.22$ with $T=0.1\Lambda^2/2m$ and $2m=1$. (Red) squares: $(T,g,m,\Lambda)$, (black) dots: $(T/2,g,m,\Lambda)$, (orange) diamonds: $(T,2g,m/2,\Lambda)$. The lower symbols are obtained for $\tilde g(T)=5$. (Blue) triangles: $(T,g,m,\Lambda)$, (green) triangles: $(T/2,g,m,\Lambda/\sqrt{2})$, (purple) triangles: $(2T,2g,m/2,\Lambda)$. The solid lines are guides to the eyes. $(\mu/T)_{\rm BKT}$ at the BKT transition is given in table~\ref{table_Tkt}.}
\label{fig_scaling}
\end{figure}

We first discuss the $x$ dependence of $\calF(x,y)$ for fixed $y$. Figure~\ref{fig_scaling} shows the phase-space pressure $\calP=\calF$ [Eqs.~(\ref{calPD})] as a function of $\mu/T$ for $\tilde g(T)=0.22$ and $\tilde g(T)=5$. We can verify that the scaling form~(\ref{pressure0}) holds by computing $\calP$ for various sets of parameters $(T,g,m,\Lambda)$. For the case $\tilde g(T)=0.22$, we choose the value $\tilde g=0.22$ for the bare interaction constant, so that the system is in the weak-coupling limit and $\tilde g(T)\simeq 0.22$ nearly temperature independent (see the discussion below). We find that the three sets of parameters $(T,g,m,\Lambda)$, $(T/2,g,m,\Lambda)$ and $(T,2g,m/2,\Lambda)$ (with $T=0.1\Lambda^2/2m$ and $2m=1$) yield the same results for the phase-space pressure $\calP$ in agreement with the expected scale invariance at weak coupling: $\calF(\mu/T,\tilde g(T))=\calF(\mu/T,\tilde g)$. In the case $\tilde g(T)=5$, the results obtained for the three sets of parameters $(T,g,m,\Lambda)$, $(
T/2,g,m,\Lambda/\sqrt{2})$ and $(2T,2g,m/2,\Lambda)$ also collapse on a single curve corresponding to the scaling function $\calF(x,y)$ (with $y\equiv \tilde g(T)$ fixed). In this case one must change simultaneously at least two parameters to keep $\tilde g(T)$ unchanged. In table~\ref{table_Tkt} we indicate the value of $(\mu/T)_{\rm BKT}$ at the BKT transition for various values of $\tilde g(T)\lesssim 1$ as obtained from the NPRG (Sec.~\ref{sec_bkt}) and Monte Carlo simulations~\cite{Prokofev02}. Note that neither our method nor the Monte Carlo simulations gives a reliable estimate of $(\mu/T)_{\rm BKT}$ in the strong-coupling limit $\tilde g(T)\gtrsim 1$. 

\begin{table}
\renewcommand{\arraystretch}{1.5}
\begin{center}
\begin{tabular}{ccccc}
\hline \hline
$\tilde g(T)$ & 0.1 & 0.22 & 0.5 & 1 \\
\hline
$(\mu/T)_{\rm BKT} $ & 0.08 & 0.15 & 0.29 & 0.51 \\
\hline
$(\mu/T)_{\rm BKT}$ (Ref.~\cite{Prokofev02}) & 0.09 & 0.17 & 0.31 & 0.51  \\
\hline \hline
\end{tabular}
\end{center}
\caption{$(\mu/T)_{\rm BKT}$ at the BKT transition for various various values of $\tilde g(T)\lesssim 1$ as obtained from the NPRG (Sec.~\ref{sec_bkt}) and Monte Carlo simulations~\cite{Prokofev02}.}
\label{table_Tkt}
\end{table}

\begin{figure}
\centerline{\includegraphics[width=6.5cm]{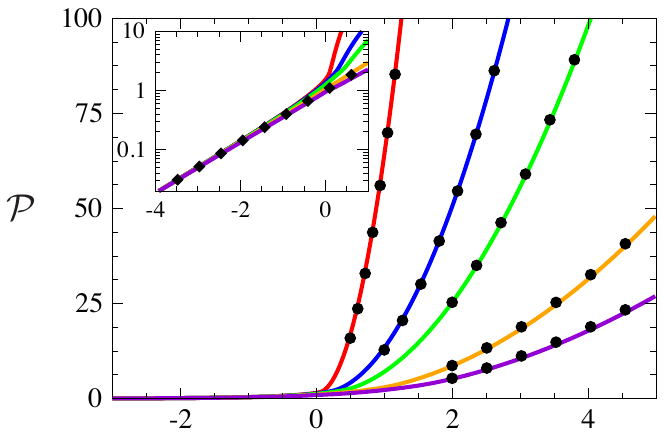}}
\centerline{\includegraphics[width=6.5cm]{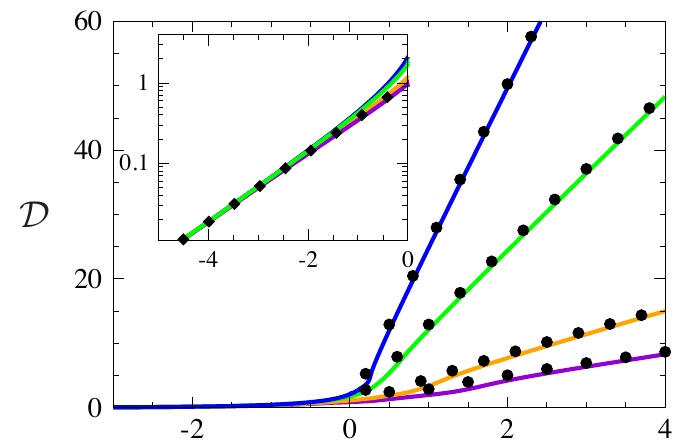}}
\centerline{\hspace{0.35cm}\includegraphics[width=6.15cm]{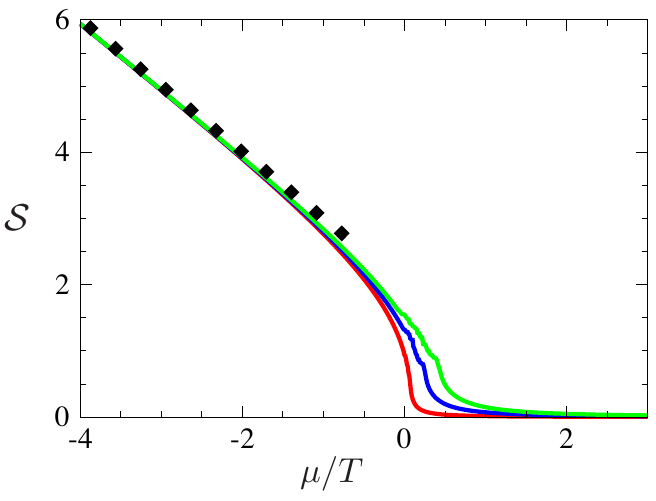}}
\caption{(Color online) Phase-space pressure $\calP(\mu,T)$, phase-space density $\calD(\mu,T)$ and entropy per particle $\calS(\mu,T)$ vs $\mu/T$.  From top to bottom, the curves correspond to $\tilde g(T)=0.1,0.5,1,3,5$ for $\calP$, $\tilde g(T)=0.5,1,3,5$ for $\calD$, and $\tilde g(T)=0.1,0.5,1$ for $\calS$. The dots show the limiting behavior~(\ref{MF}) valid for $\mu\gg T$ and the diamonds the classical gas result~(\ref{classical}). [For numerical reasons, it is difficult to compute the entropy when $\tilde g(T)\gtrsim 1$.]}
\label{fig_F1} 
\end{figure}

Figure~\ref{fig_F1} shows the phase-space pressure $\calP=\calF$, the phase-space density $\calD=\calF^{(1,0)}$ and the entropy per particle $\calS(\mu,T)$ as a function of $\mu/T$ for various values of $\tilde g(T)$ [Eqs.~(\ref{calPD}-\ref{calS})]. At large and negative chemical potential ($|\mu|/T\gg 1$), we find that the system behaves as a classical dilute gas,
\begin{equation}
\begin{gathered}
\calP(\mu,T) = \calD(\mu,T) = e^{-|\mu|/T} , \\
\calS(\mu,T) = 2 - \frac{\mu}{T} ,
\end{gathered}
\label{classical}
\end{equation}
which implies 
\begin{equation}
\calF(x,y)=e^x \quad \mbox{for} \quad  x<0 \quad  \mbox{and} \quad |x|\gg 1. 
\label{classical1}
\end{equation}
In the opposite limit of a large positive chemical potential ($\mu/T\gg 1$), the pressure can be approximated by its zero-temperature limit. Using the expression~(\ref{P2}), we obtain 
\begin{equation}
\begin{split}
\calP(\mu,T) &=  \left(\frac{\mu}{T}\right)^2 \left( \frac{2\pi}{\tilde g(\mu)} + \quarter \ln 2 - \frac{1}{8} \right), \\
\calD(\mu,T) &= \frac{\mu}{T} \left( \frac{4\pi}{\tilde g(\mu)} + \half \ln 2 - \frac{1}{2} \right) , \\
\calS(\mu,T) &= 0 ,
\end{split}
\label{MF}
\end{equation}
{\it i.e.} 
\begin{equation}
\calG(x,y)=\left( \frac{2\pi}{y} + \quarter \ln 2 - \frac{1}{8} \right) \quad \mbox{for} \quad x\ll 1 , 
\label{MF1}
\end{equation}
where $\calG\equiv\calG_2$ is the scaling function defined in Eq.~(\ref{calG}) ($x\equiv T/\mu$)~\cite{note5}. Without the additive constants $-1/8+(\ln 2)/4$ and $-1/2+(\ln2)/2$, Eqs.~(\ref{MF}) coincide with the mean-field result assuming an effective interaction constant $g(\mu)$. These constants can be omitted in the weak-coupling limit $\tilde g(\mu)\ll 1$. As pointed out in Refs.~\cite{Prokofev01,Prokofev02}, in the  weak-coupling limit -- where the BKT transition temperature $\Tkt$ can be easily determined (see Sec.~\ref{sec_bkt}) -- the approximation~(\ref{MF1}) remains remarkably accurate all the way down to the transition point $(\mu/T)_{\rm BKT}$. We also observe that the limiting behaviors~(\ref{classical},\ref{classical1}) and (\ref{MF},\ref{MF1}) are very well satisfied not only in the weak-coupling limit~\cite{Prokofev01,Prokofev02} but also in the strong-coupling limit where $\tilde g(T)\gtrsim 1$ (see Fig.~\ref{fig_F1}). 

The crossover regime $|\mu|\ll T$ is more difficult to analyze in simple terms, and a full numerical solution of the RG equations is necessary (see however Sec.~\ref{subsec_Fy} and Appendix~\ref{app_muzero} for an analytical solution in the case $\mu=0$). 

In the weak-coupling limit $\tilde g=2mg\ll 1$, the scattering length $a_2$ is exponentially small. This implies that the renormalized interaction constant $\tilde g(T) \simeq \tilde g$ is nearly temperature independent except for exponentially small temperatures (which are experimentally unreachable)~\cite{note10}. It follows that the phase-space pressure and density and the entropy per particle, 
\begin{equation}
\begin{split}
\calP(\mu,T) &= \calF\left(\frac{\mu}{T},\tilde g\right), \\ 
\calD(\mu,T) &= \calF^{(1,0)}\left(\frac{\mu}{T},\tilde g\right), \\
\calS(\mu,T) &= 2 \frac{\calP(\mu,T)}{\calD(\mu,T)} - \frac{\mu}{T} ,
\end{split}
\label{PDweak}
\end{equation}
can be considered as functions of $\mu/T$ only, with the microscopic interaction constant $\tilde g$ entering the scaling function $\calF$ as a parameter. The equation of state then exhibits an approximate scale invariance (with no characteristic energy scales other than $\mu$ and $T$)~\cite{Prokofev01,Prokofev02}.

\subsection{$\calF(0,y)$ vs $y$}
\label{subsec_Fy}

\begin{figure}
\centerline{\hspace{0.2cm}\includegraphics[width=6.5cm]{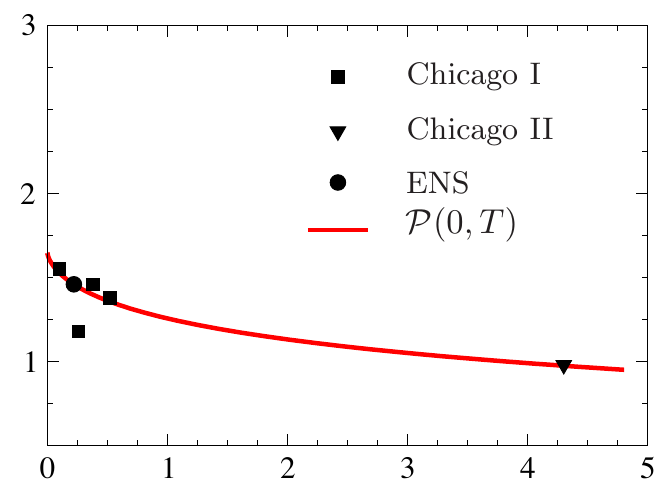}}
\centerline{\hspace{0.2cm}\includegraphics[width=6.5cm]{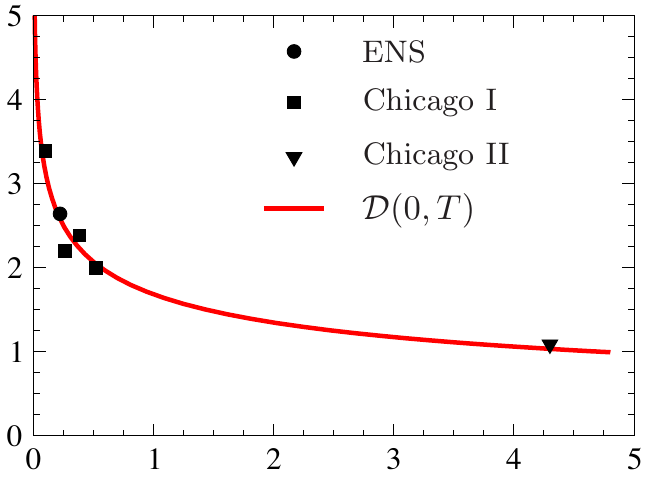}}
\centerline{\includegraphics[width=6.75cm]{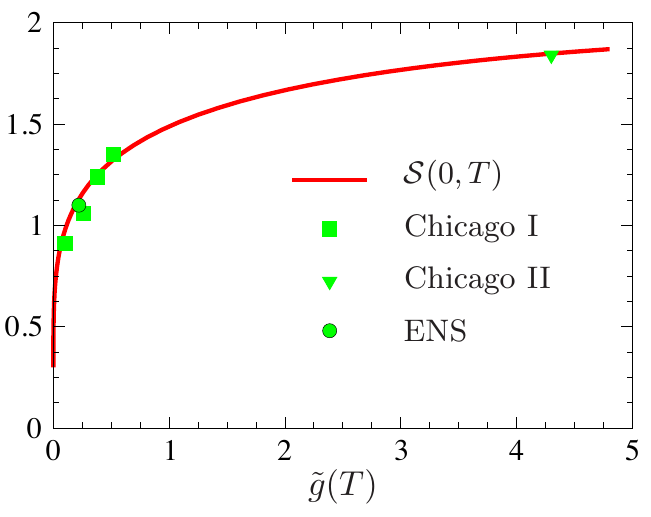}}
\caption{(Color online) Phase-space pressure $\calP(0,T)$, phase-space density $\calD(0,T)$ and entropy per particle $\calS(0,T)$ vs $\tilde g(T)$. The symbols show the the data of the ENS, Chicago I and Chicago II (with $\mu=\mu_c$) experiments~\cite{Yefsah11,Hung11,Zhang12} (see Sec.~\ref{sec_exp}).}
\label{fig_F2}
\end{figure}

The limit $|x|\ll 1$ is particularly interesting as it corresponds to the quantum critical regime. In this section, we discuss the function $\calF(0,y)$. Figure~\ref{fig_F2} shows $\calP(0,T)$, $\calD(0,T)$ and $\calS(0,T)$ as a function of $\tilde g(T)$. We show in Appendix~\ref{app_muzero} that  
\begin{equation}
\calP(0,T) \equiv \calF(0,\tilde g(T)) \simeq \frac{\pi^2}{6} -  \frac{\tilde g(T)}{2\pi} \ln^2\left(\frac{2\pi}{\tilde g(T)}\right)  
\end{equation}
for $\tilde g(T)\to 0$. The result $\lim_{T\to 0}\calP(0,T)=\pi^2/6$ is exact. Experimentally, however, this limiting behavior cannot be observed due to the logarithmic temperature dependence of $\tilde g(T)$. In the weak-coupling limit, $\tilde g(T)=\tilde g$ is nearly temperature independent, and the phase-space pressure takes the form 
\begin{equation}
\calP(0,T) = \calF(0,\tilde g) ,
\end{equation}
where $\calF(0,\tilde g)\leq \lim_{y\to 0}\calF(0,y)=\pi^2/6$. In the strong-coupling limit, $\calP(0,T)$ exhibits a  weak temperature dependence coming from that of $\tilde g(T)$, but again reaching the limiting value $\lim_{T\to 0}\calP(0,T)=\pi^2/6$ requires extremely small (unrealistic) temperatures.

\subsection{Thermodynamics of the Bose-Hubbard model}
\label{subsec_bh}

\begin{figure}
\centerline{\includegraphics[width=7.3cm]{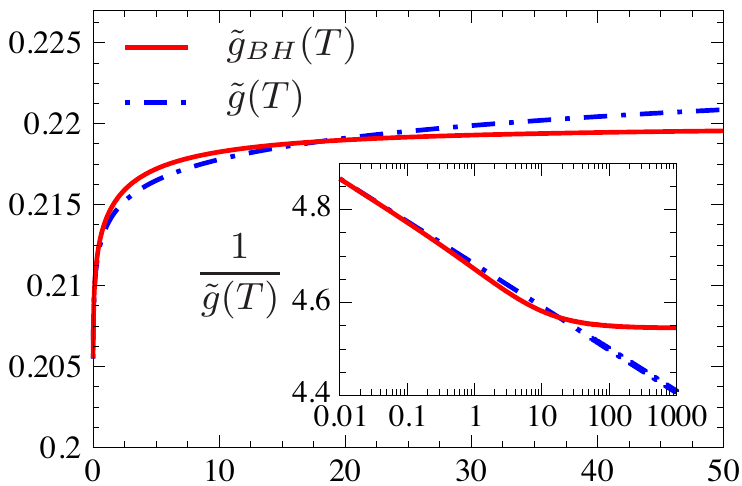}}
\centerline{\hspace{0.435cm}\includegraphics[width=6.735cm]{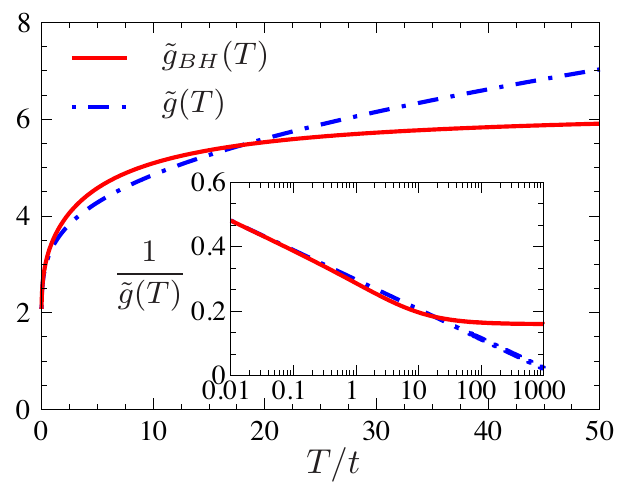}}
\caption{(Color online) (Red) solid lines: dimensionless interaction constant $\tilde g_{\rm BH}(T)$ in the two-dimensional Bose-Hubbard model for $U/t=0.22$ (top) and $U/t=6.25$ (bottom) (note the different scales on the vertical axes). The (blue) dash-dotted lines show the universal limit $\tilde g(T)$.}
\label{fig_gtlat}
\end{figure}
\begin{figure}
\centerline{\includegraphics[width=7cm]{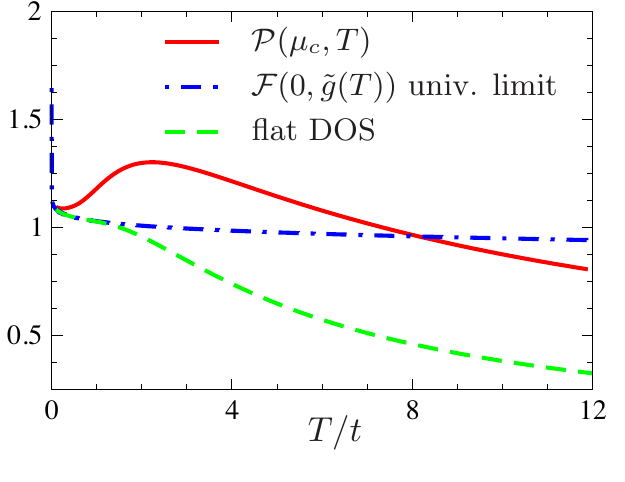}}
\caption{(Color online) Phase-space pressure $\calP(\mu_c,T)$ vs $T/t$ for $U/t=6.25$. The (green) dashed line shows the result obtained for a flat density of states (DOS) in the energy window $[0,8t]$. The (blue) dash-dotted line shows the universal limit $\calF(0,\tilde g(T))$.}
\label{fig_P_lat}
\centerline{\includegraphics[width=7cm]{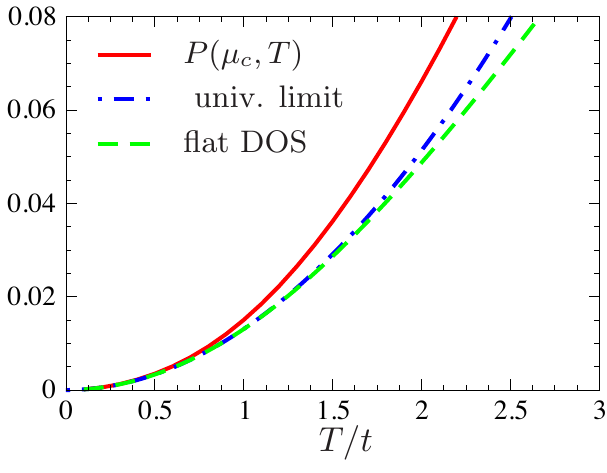}}
\caption{(Color online) Pressure $P(\mu_c,T)$ vs $T/t$ for $U/t=6.25$. The (green) dashed line shows the result obtained for a flat density of states in the energy window $[0,8t]$. The (blue) dash-dotted line shows the universal limit $T^2/(4\pi tl^2)\calF(0,\tilde g(T))$.}
\label{fig_Ptot_lat}
\centerline{\includegraphics[width=7cm]{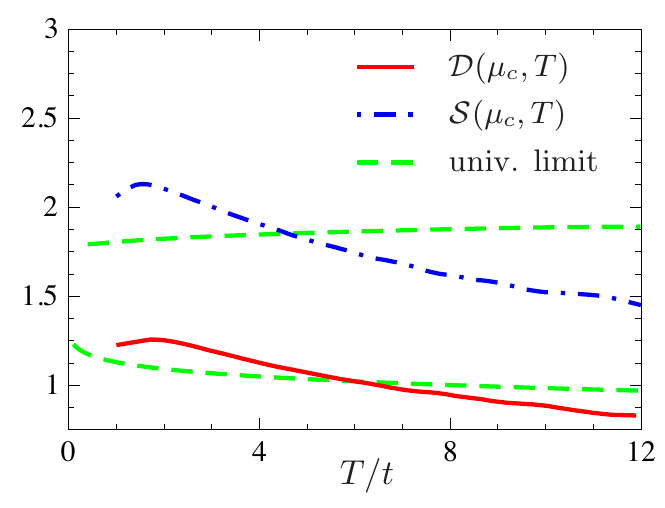}}
\caption{(Color online) Phase-space density $\calD(\mu_c,T)$ ((red) solid line) and entropy per particle $\calS(\mu_c,T)$ ((blue) dash-dotted line) vs $T/t$ for $U/t=6.25$. The (green) dashed lines show the corresponding universal limits.}
\label{fig_DS_lat}
\end{figure}
\begin{figure}
\centerline{\includegraphics[width=7cm]{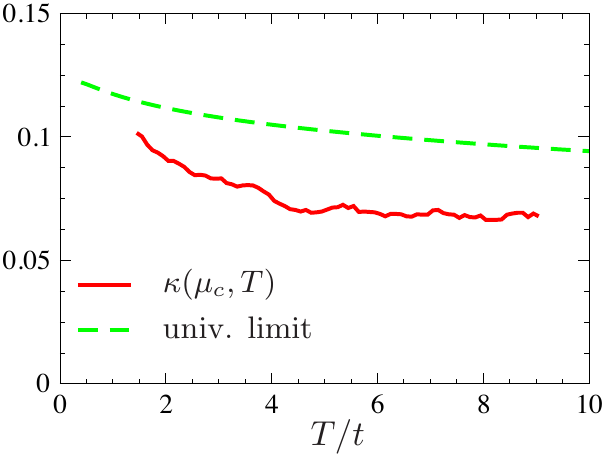}}
\caption{(Color online) Compressibility $\kappa(\mu_c,T)$ vs $T/t$ for $U/t=6.25$. The (green) dashed line shows the corresponding universal limit. (The numerical noise in the NPRG result follows from taking the second-order derivative of the pressure with respect to $\mu$.)}
\label{fig_kappa}
\end{figure}

In this section we discuss the results obtained in the two-dimensional Bose-Hubbard model [Eq.~(\ref{BH})] using the lattice version of the NPRG~\cite{Rancon11a,Rancon11b}. The energy-dependent interaction constant is defined by 
\begin{equation}
\gbh(\eps) = \frac{U}{1+U\Pi(\eps)} , 
\label{gbh}
\end{equation}
with
\begin{equation}
\Pi(\eps) = l^2 \int \frac{d^2q}{(2\pi)^2} \frac{1}{2(\eps_\q+\eps)} ,
\label{pibh}
\end{equation}
where $\eps_\q$ is the lattice dispersion of the boson~[Eq.~(\ref{epsq})]. This definition, which is also that used in Ref.~\cite{Zhang12}, is justified in Appendix~\ref{app_disp}. In the low-energy limit $\eps\to 0$, it coincides with the universal form $l^{-2}g(\eps)$ [Eq.~(\ref{gT})], obtained from the continuum model with boson mass $m=1/2tl^2$  and scattering length $a_2$ given by Eq.~(\ref{a2lat}). The dimensionless interaction constant $\tilde g_{\rm BH}(T)=2ml^2\gbh(T)=\gbh(T)/t$ is shown in Fig.~\ref{fig_gtlat} for $U/t=0.22$ and $U/t=6.25$. In both cases, $\tilde g_{\rm BH}(T)$ is well approximated by its universal limit $\tilde g(T)$ [Eq.~(\ref{gT})] for $T\lesssim 8t$ (see insets in Fig.~\ref{fig_gtlat}). 

\begin{figure}
\centerline{\includegraphics[width=7.13cm]{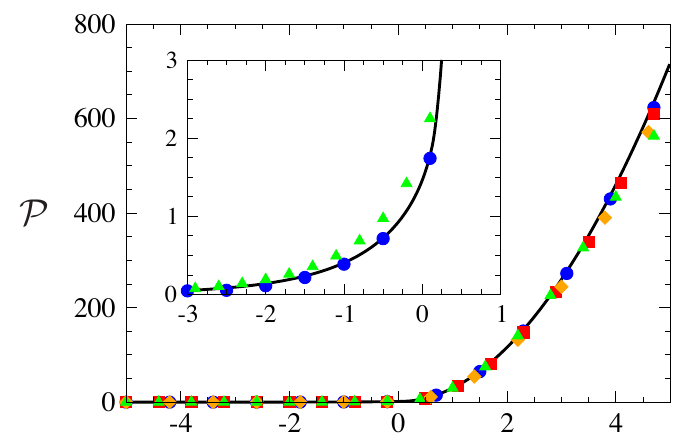}}
\centerline{\hspace{0.155cm}\includegraphics[width=7cm]{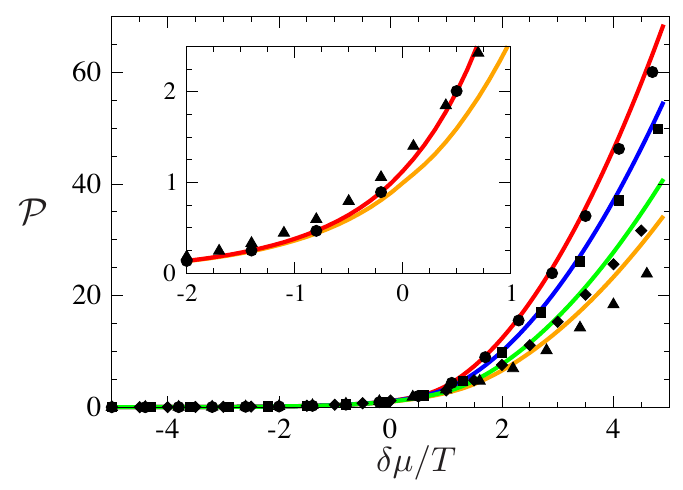}}
\caption{(Color online) Phase-space pressure $\calP(\mu,T)$ vs $\delta\mu/T$ for $U/t=0.22$ (top) and $U/t=6.25$ (bottom). $T/t=1/100$ (circles), $1/10$ (squares), $1$ (diamonds) and $3$ (triangles). The solid lines correspond to $\calF(\delta\mu/T,\tilde g(T))$. For $t/U=0.22$, $\mu/T\simeq 0.15$ at the BKT transition (see table~\ref{table_Tkt} and Sec.~\ref{sec_bkt}).}
\label{fig_scaling_lat} 
\centerline{\includegraphics[width=6.5cm]{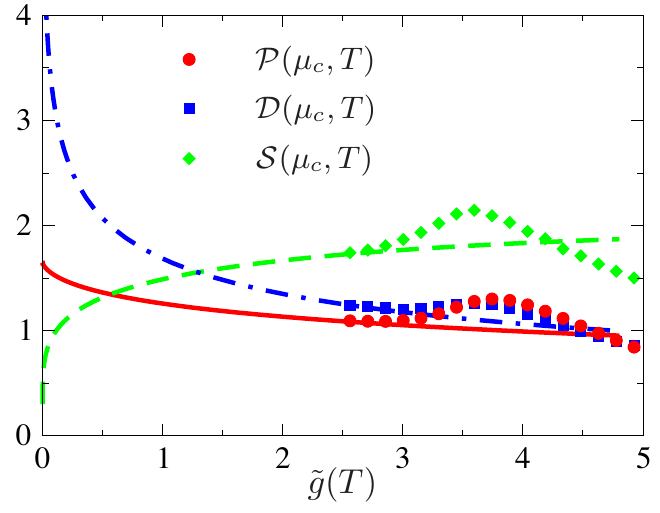}}
\caption{(Color online) Phase-space pressure $\calP(\mu_c,T)$, phase-space density $\calD(\mu_c,T)$, and entropy per particle $\calS(\mu_c,T)$ vs $\tilde g(T)$ ($U/t=6.25$). The lines show the universal limit obtained from the scaling function $\calF(0,\tilde g(T))$ and its derivatives.}
\label{fig_PDS_scaling}
\end{figure}

Figure~\ref{fig_P_lat} shows the phase-space pressure $\calP(\mu_c,T)$ vs $T/t$ for $U/t=6.25$ and $\mu=\mu_c=-4t$. We observe a maximum around $T/t\sim 2.5$ due to the enhanced density of states of the square lattice near the band center~\cite{note11}. This maximum disappears if we consider a flat density of states in the energy window $[0,8t]$. Comparing $\calP(\mu_c,T)$ and $\calF(0,\tilde g(T))$ (with $\tilde g(T)$ the universal limit of $\tilde g_{\rm BH}(T)$ discussed above) we see that the universal limit, where $\calP(\mu_c,T)$ becomes a universal function of $ma_2^2T$, is reached only at very low temperatures $T\ll t$. The identification of $t$ as the crossover temperature scale for quantum critical behavior is confirmed by the $T$ dependence of the pressure. For $T\lesssim t$, one finds that $P(\mu_c,T)=T^2/(4\pi tl^2)\calP(\mu_c,T)$ is well approximated by the universal limit $T^2/(4\pi tl^2)\calF(0,\tilde g(T))$ (Fig.~\ref{fig_Ptot_lat}). The phase-space density $\calD(\mu_c,T)$ and entropy per 
particle $\calS(\mu_c,T)$ are shown in Fig.~\ref{fig_DS_lat} (the low-temperature regime where $\calD(\mu_c,T)$ and $\calS(\mu_c,T)$ coincide with their universal limits is not shown). 

The fact that the universal regime is reached only at low temperatures can also be seen in the temperature dependence of the compressibility $\kappa=\partial^2 P/\partial\mu^2$ (Fig.~\ref{fig_kappa}). Although it is difficult to numerically compute the second-order derivative of the pressure with respect to $\mu$, our results clearly show that $\kappa(\mu_c,T)$ is below the universal limit $(1/4\pi tl^2)\calF^{(2,0)}(0,\tilde g(T))$. We also note that while $\kappa(\mu_c,T)$ varies weakly with $T$ in the temperature range $[t,10t]$, it should eventually diverge as $T\to 0$ (see Eq.~(\ref{rgeq11}) in appendix~\ref{app_muzero}). We thus disagree with the conclusion of Ref.~\cite{Fang11} that quantum criticality is observed below a characteristic temperature of the order of the single-particle bandwidth $8t$~\cite{note17}. 

Figure~\ref{fig_scaling_lat} shows the phase-space pressure $\calP(\mu,T)$ versus $\delta\mu/T$ for $U/t=0.22$ and $U/t=6.25$ ($\delta\mu=\mu-\mu_c$), and various temperatures ranging from $t/100$ to $3t$. In the weak-coupling limit $U/t=0.22$, $\tilde g_{\rm BH}(T)=U/t$ is nearly temperature independent in the temperature range $[t/100,3t]$ (see Fig.~\ref{fig_gtlat}). At very low temperatures, we  obtain a perfect agreement between $\calP(\mu,T)$ and the universal scaling function $\calF(\delta\mu/T,\tilde g)$ (with $\tilde g=0.22$). At higher temperatures, when $T\sim t$, we observe that $\calP(\mu,T)$ slightly deviates from $\calF(\delta\mu/T,\tilde g)$, in particular for large values of $\mu$. This agrees with the previous observation that $\calP(\mu_c,T)$ reaches the universal limit only for $T\lesssim t$ (see Figs.~\ref{fig_P_lat} and \ref{fig_Ptot_lat}). For $U/t=6.25$, we again find a good agreement between $\calP(\mu,T)$ and $\calF(\delta\mu/T,\tilde g(T))$ at low temperatures and small chemical 
potential $\delta\mu$, but deviations are clearly visible at higher temperatures or larger values of $\delta\mu/T$.

The phase-space pressure $\calP(\mu_c,T)$, phase-space density $\calD(\mu_c,T)$ and entropy per particle $\calS(\mu_c,T)$ vs $\tilde g(T)$ are shown in Fig.~\ref{fig_PDS_scaling} for $U/t=6.25$. The maximum around $\tilde g(T)\sim 3.5$ is due to the enhanced density of states of the square lattice near the band center~\cite{note11}. For $\tilde g(T)\lesssim 2.5$, we recover the universal limit where the thermodynamics is determined by the scaling function $\calF(\delta\mu/T,\tilde g(T))$.

\section{Comparison with experiments} 
\label{sec_exp}

\begin{figure}
\centerline{\includegraphics[width=6.6cm]{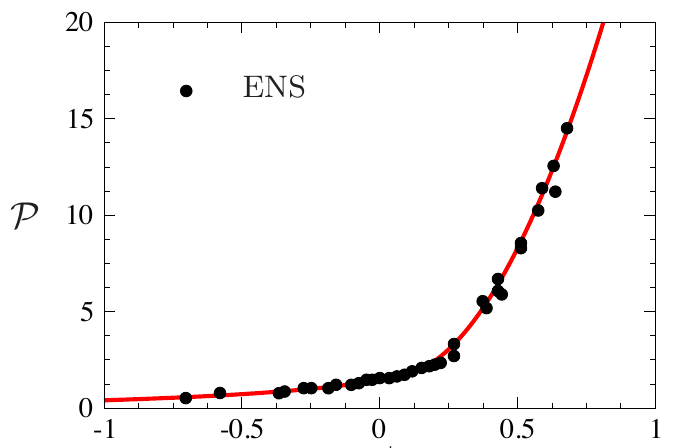}}
\centerline{\hspace{0.05cm}\includegraphics[width=6.5cm]{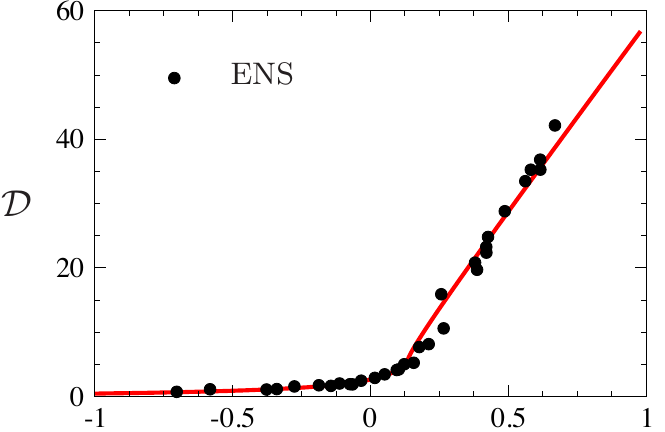}}
\centerline{\hspace{0.1cm}\includegraphics[width=6.37cm]{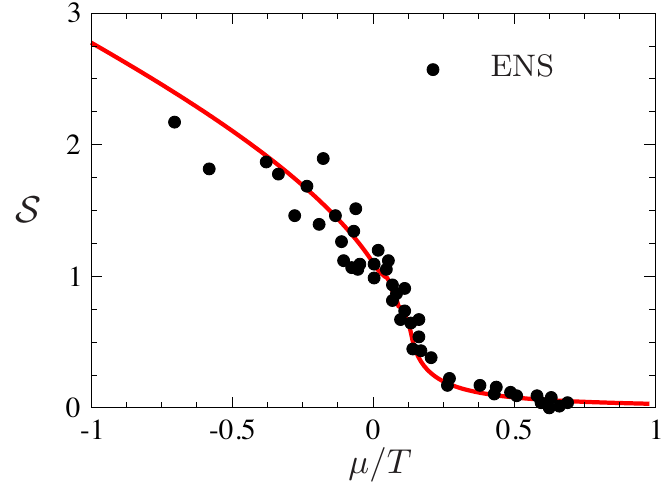}}
\caption{(Color online) Phase-space pressure $\calP(\mu,T)$, phase-space density $\calD(\mu,T)$ and entropy per particle $\calS(\mu,T)$ vs $\mu/T$ in the ENS experiment~\cite{Yefsah11}. The (red) solid lines show the NPRG results.} 
\label{fig_ens} 
\end{figure}

In this section, we compare our theoretical results for the scaling function $\calF\equiv\calF_2$ with three recent experiments on two-dimensional Bose gases. The first experiment was realized with a gas of $^{87}$Rb atoms with scattering length $a_3=5.3$\,nm and a thickness $l_z=240$\,nm in the confining direction leading to a dimensionless interaction constant $\tilde g=2mg=0.22$~\cite{Yefsah11,note6}. The second one was performed with $^{133}$Cs atoms and a scattering length $a_3$ controlled by a Feshbach resonance and varying in the range $2-10$\,nm resulting in $\tilde g=0.1-0.52$~\cite{Hung11}. The last one was realized with a $^{133}$Cs atom gas in an optical lattice and can be described by the Bose-Hubbard model with $t=2.7$\,nK, $U=16.7$\,nK ({\it i.e.} $U/t=6.25$), and a temperature varying in the range $5.8-32$\,nK ({\it i.e.} $2.15t-32t$)~\cite{Zhang12,note7}. This leads to a temperature-dependent dimensionless interaction constant $\tilde g_{\rm BH}(T)$ varying between $3.95$ and $5.75$. We 
refer to these experiments as the ``ENS", ``Chicago I" and ``Chicago II" experiments, respectively.  

In Fig.~\ref{fig_ens}, we compare the NPRG results with the ENS experiment. For $\tilde g=0.22$, the temperature dependence of $\tilde g(T)$ is negligible so that we expect the scaling forms~(\ref{PDweak}), which express $\calP$, $\calD$ and $\calS$ as universal functions of $\mu/T$ and $\tilde g$, to be very well satisfied. We find a nearly perfect agreement between the experimental data and the NPRG calculation of the universal function $\calF(\mu/T,\tilde g)$ (without any fitting parameter). 

\begin{figure}
\centerline{\includegraphics[width=7.25cm]{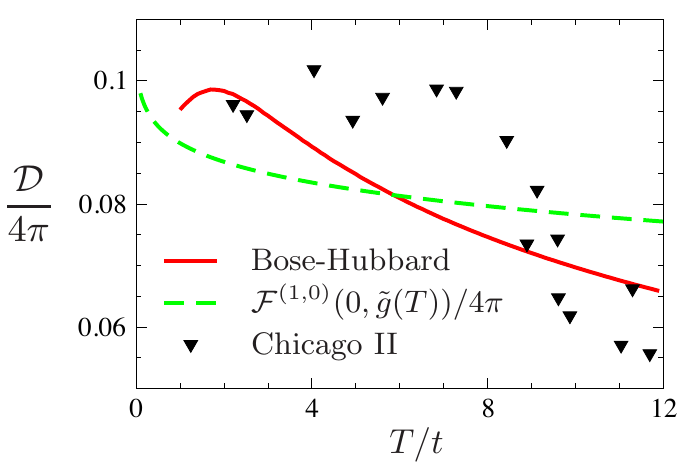}}
\caption{(Color online) Temperature dependence of the phase-space density $\calD(\mu_c,T)$ in the Chicago II experiment (triangles)~\cite{Zhang12}. The (red) solid line shows the NPRG result and the (green) dashed one the universal limit.} 
\label{fig_chin3}
\end{figure}
\begin{figure}
\centerline{\includegraphics[width=4.4cm]{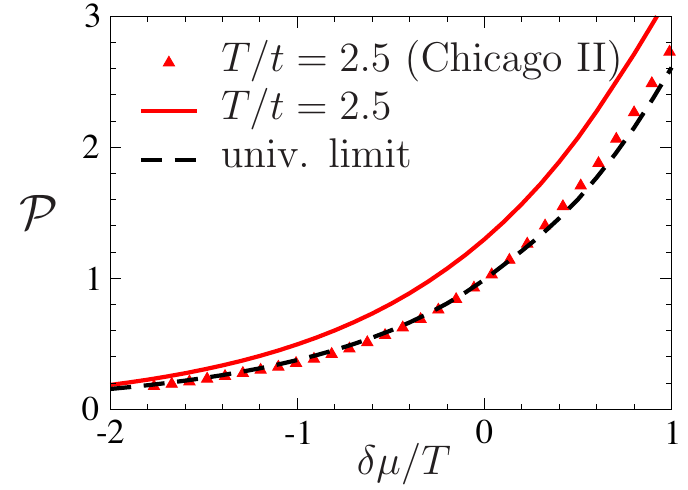}
\includegraphics[width=3.9cm]{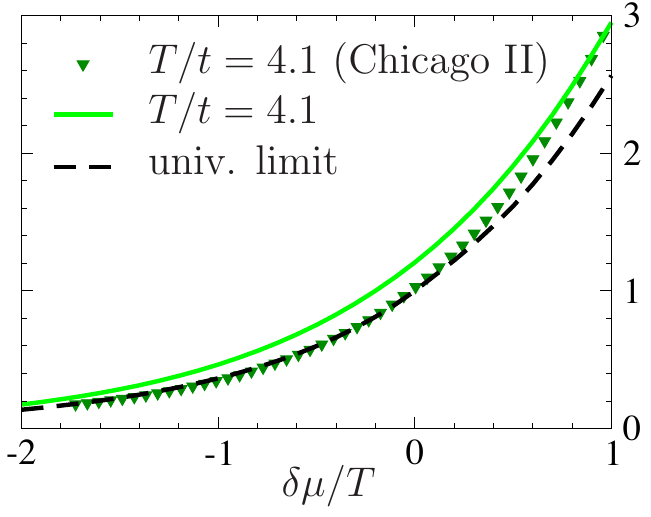}}
\caption{(Color online) Phase-space pressure $\calP(\mu,T)$ vs $\delta\mu/T$ for $T=6.7\,$nK and $T=11\,$nK in the Chicago II experiment~\cite{Zhang12}. The solid lines show the NPRG results obtained in the Bose-Hubbard model and the dashed lines the universal limit.}
\label{fig_chin24a}
\centerline{\includegraphics[width=7.6cm]{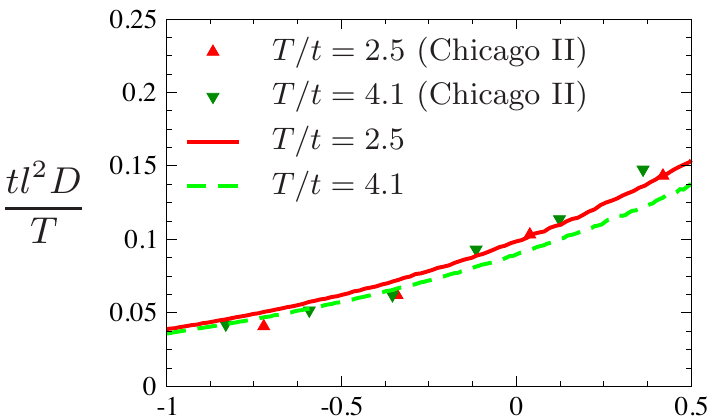}}
\centerline{\hspace{0.8cm}\includegraphics[width=6.55cm]{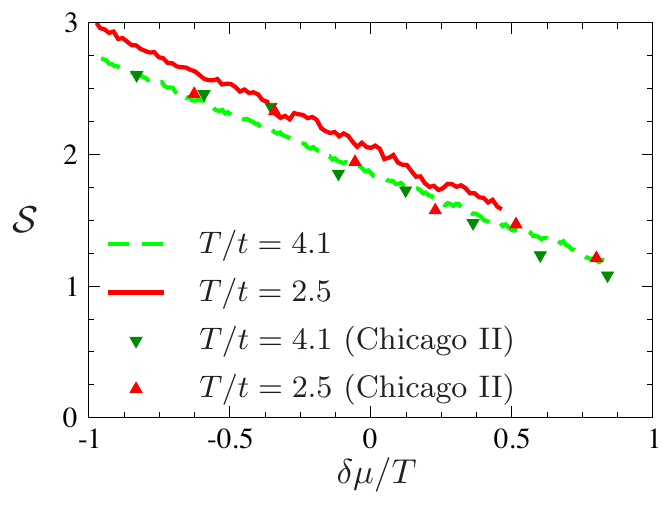}}
\caption{(Color online) Phase-space density $\calD(\mu,T)$ and entropy per particle $\calS(\mu,T)$ vs $\delta\mu/T$ for $T=6.7$ and $T=11\,$nK in the Chicago II experiment~\cite{Zhang12}. The solid and dashed lines show the NPRG results obtained in the Bose-Hubbard model.} 
\label{fig_chin24b}
\end{figure}

In Sec.~\ref{subsec_bh}, we have shown that a Bose gas in an optical lattice, described by the Bose-Hubbard model with $U/t=6.25$, reaches the universal limit only at temperatures of the order of $t$. In the Chicago II experiment, the lowest temperature $T\sim 2.15t$ is above $t$, and we should therefore expect experimental data to agree only approximately with results obtained from the universal function $\calF$. The temperature dependence of the phase-space density $\calD(\mu_c,T)$ is shown in Fig.~\ref{fig_chin3}. There is an overall agreement between the experimental data and the NPRG results but the existence of a plateau for $T\lesssim 8t$ followed by a strong suppression of $\calD(\mu_c,T)$ at higher temperatures, as advocated in Ref.~\cite{Zhang12}, is not supported by the theory. In Fig.~\ref{fig_chin24a} we show the phase-space pressure $\calP(\mu,T)$ vs $\delta\mu/T$ for $T/t=2.5$ and $T/t=4.1$. As expected the NPRG results show deviations from the universal limit $\calF(\delta\mu,\tilde g(T))$ (
note that $\tilde g(T)$ is nearly temperature independent in the temperature range $[2.5t-4.1t]$). For large and negative chemical potential $\delta\mu$, the pressure is very well approximated by the classical dilute gas expression
\begin{equation}
\calP(\mu,T) = 4\pi te^{-|\delta\mu|/T} l^2\int_\q e^{-\eps_\q/T} .
\label{calP1} 
\end{equation}
The difference with the universal limit $\calP=e^{-|\delta\mu|/T}$ (Sec.~\ref{subsec_Fx}) is entirely due to the difference between the lattice dispersion $\eps_\q$ [Eq.~(\ref{epsq})] and the free quadratic  dispersion $\q^2/2m$ with $m=1/2tl^2$. For $T/t=4.1$, Eq.~(\ref{calP1}) gives $\calP\simeq 1.3 e^{-|\delta\mu|/T}$ when $\delta\mu/T\lesssim -2$. On the other hand the experimental data show a remarkable agreement between the phase space pressure $\calP$ and the universal scaling function $\calF$ with only a small difference for positive $\delta\mu$. Such an agreement is difficult to understand in the framework of the Bose-Hubbard model. In particular, one would expect $\calP$ to differ from $e^{-|\delta\mu|/T}$ for large and negative $\delta\mu$ and $T/t\simeq 2-4$ due to lattice effects (see the discussion above). The phase-space density $\calD(\mu,T)$ and the entropy per particle $\calS(\mu,T)$ vs $\delta\mu/T$ for $T/t=2.5$ and $T/t=4.1$ are shown in Fig~\ref{fig_chin24b}; there is a good agreement 
between theory and experiment. 

The ENS, Chicago I and Chicago II experiments can be used to obtain $\calP(\mu_c,T)$, $\calD(\mu_c,T)$ and $\calS(\mu_c,T)$ as a function of the effective interaction constant $\tilde g(T)$. The results are shown in Fig.~\ref{fig_F2}. For all three experiments, we obtain a very good agreement with the universal limit~(\ref{calPD},\ref{calS}). This confirms that both the ENS and Chicago I experiments deal with a weakly interacting Bose gas in the universal regime. As for the Chicago II experiment, such a good agreement is partially accidental since for $\tilde g_{\rm BH}(T)\simeq\tilde g(T) \simeq 4.3$ (the relevant value of $\tilde g_{\rm BH}(T)$ corresponding to the experimental data shown in Fig.~\ref{fig_F2}), $\calS(\mu_c,T)$ turns out to be very close to the universal limit even though the system has not reached the universal regime yet (see Fig.~\ref{fig_PDS_scaling}). We also note that for this value of this interaction constant, $\calP(\mu_c,T)$ and $\calD(\mu_c,T)$ are nearly equal, which implies 
that $P(\mu_c,T)\simeq T D(\mu_c,T)$ as observed in the Chicago II experiment. 

\section{BKT transition temperature} 
\label{sec_bkt}

In this section we show how the BKT transition temperature $\Tkt$ can be estimated from the NPRG approach. For the classical O(2) model, the NPRG reproduces most of the universal properties of the BKT transition~\cite{Graeter95,Gersdorff01}. In particular one finds a value $\tilde\rho_0^*$ of the dimensionless order parameter (the spin-wave ``stiffness'') such that the beta function $\beta( \tilde\rho_{0,k})=k\dk \tilde\rho_{0,k}$ nearly vanishes for $\tilde\rho_{0,k}\geq \tilde\rho_0^*$ (here $k$ denotes the RG momentum scale, see Appendix~\ref{app_nprg}). This implies the existence of a line of quasi-fixed points and enables to identify a low-temperature phase ($T<\Tkt$) where the running of the stiffness $\tilde\rho_{0,k}$, after a transient regime, becomes very slow, implying a very large (although not strictly infinite as expected in the low-temperature phase of the BKT transition) correlation length $\xi$. In this low-temperature phase, the anomalous dimension $\eta_k$ depends on the (slowly varying) 
stiffness $\tilde\rho_{0,k}$. It takes its largest value $\sim 1/4$ when the RG flow crosses over to the disordered (long-distance) regime (for $\tilde\rho_{0,k}\sim\tilde\rho_0^*$ and $k\sim\xi^{-1}$), and is then rapidly suppressed as $\tilde\rho_{0,k}$ further decreases. On the other hand, the beta function is well approximated by $\beta( \tilde\rho_{0,k})=\const\times(\tilde\rho_0^*-\tilde\rho_{0,k})^{3/2}$ for $\tilde\rho_{0,k}\leq\tilde\rho_0^*$, and the essential scaling $\xi\sim e^{\const/(T-\Tkt)^{1/2}}$ of the correlation length above the BKT transition temperature $\Tkt$ is reproduced~\cite{Gersdorff01}. Thus, although the NPRG approach does not yield a low-temperature phase with an infinite correlation length, it nevertheless allows us to estimate the BKT transition temperature from the value of $\tilde\rho^*_0$. A reasonable estimate of the BKT transition in the two-dimensional XY model has been obtained using the lattice NPRG~\cite{Machado10}. Here we use the NPRG to determine the BKT 
transition temperature in a two-dimensional Bose gas~\cite{note13}.

\begin{figure}
\centerline{\includegraphics[width=7.3cm]{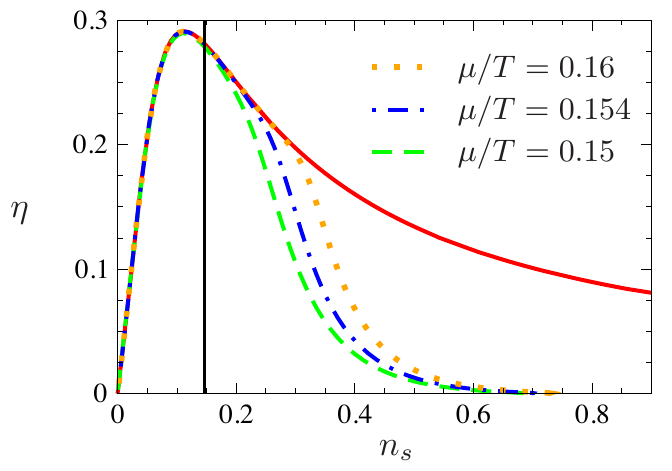}}
\caption{(Color online) Flow trajectories in the plane $(n_s,\eta)$ for a two-dimensional Bose gas with a dimensionless interaction constant $\tilde g=0.22$. The vertical line indicates the value of $n_s^*$. The (red) solid line shows the line of quasi-fixed points  for $n_s\geq n_s^*$. The critical trajectory (which joins the line of quasi-fixed points for $n_s=n_s^*$) corresponds to $\mu/T\simeq 0.154$.}
\label{fig_eta_ns}
\end{figure}

The flow trajectories in the plane $(n_s,\eta)$ are shown in Fig.~\ref{fig_eta_ns} for the continuum model with $\tilde g=0.22$. $n_s$ denotes the superfluid density and is analog to the dimensionless order parameter $\tilde\rho_0$ of the classical O(2) model. At sufficiently low temperatures, the trajectories join a line of quasi-fixed points where the RG flow is very slow, before eventually crossing over to the disordered phase ($n_s\to 0$). The value of $n_s$ at the merging point with the line of quasi-fixed points depends on the temperature and chemical potential of the Bose gas. We estimate the BKT transition temperature by the trajectory for which the merging point corresponds to the  value $n_s^*$ (analog to $\tilde\rho_0^*$ in the classical O(2) model) of the superfluid density. A precise determination of the value of $n_s^*$ (which can be obtained by fitting the beta function $k\dk n_{s,k}$ by $\const\times(n_s^*-n_{s,k})^{3/2}$ for $n_{s,k}<n_s^*$) is however difficult as it requires the full $\
calO(\partial^2)$ expansion of the effective action while we solve the NPRG equation within a simple truncation of the effective potential [Eq.~(\ref{trunc})]. Nevertheless, since the BKT transition in the Bose gas model and the classical O(2) model is controlled by the same fixed point, we expect the ratio $n_s^*/n_s^{\rm max}$, where $n_s^{\rm max}$ is the value of $n_s$ for which $\eta$ is maximum (Fig.~\ref{fig_eta_ns}), to be equal to $\tilde\rho_0^*/\tilde\rho_0^{\rm max}$. 

\begin{figure}
\centerline{\includegraphics[width=8.cm]{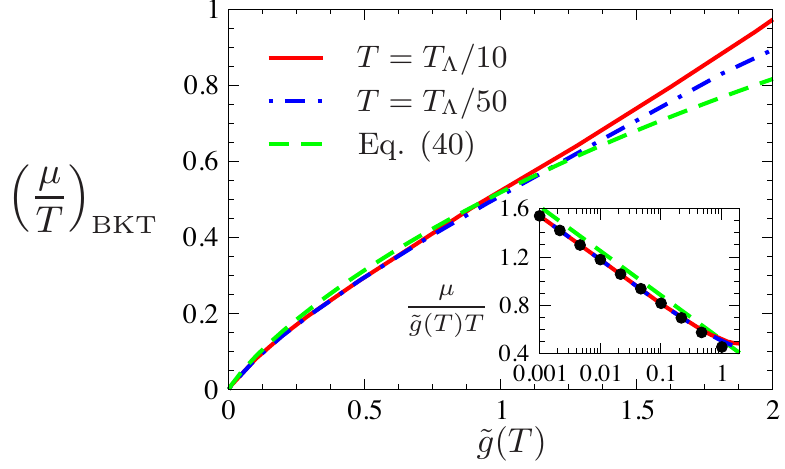}}
\caption{(Color online) Ratio $(\mu/T)_{\rm BKT}$ vs $\tilde g(T)$ for $T=T_\Lamb/10$ and $T=T_\Lamb/50$, where $T_\Lamb=\Lamb^2/2m$. The (green) dotted line corresponds to the expression~(\ref{Tktmc}) with the Monte Carlo result $\zeta=13.2$~\cite{Prokofev01,Prokofev02}.}
\label{fig_Tkt}
\end{figure}

Using this method, we have verified that the ratio $\mu/T$ at the BKT transition is a universal function of $\tilde g(T)$, {\it i.e.} 
\begin{equation}
\left( \frac{\mu}{T}\right)_{\rm BKT} = \calH\bigl(\tilde g(T)\bigr) ,
\label{ktuniv}
\end{equation}
with $\calH$ a universal function. Equivalently, since $\tilde g(T)$ is a function of $ma_2^2T$, $(\mu/T)_{\rm BKT}$ can be seen as a universal function of $ma_2^2T$ or $ma_2^2\mu$. Figure~\ref{fig_Tkt} shows $(\mu/T)_{\rm BKT}$ obtained for two different temperatures, $T=T_\Lambda/10$ and $T_\Lambda/50$ ($T_\Lambda=\Lamb^2/2m$), and a range of values of $\tilde g$. The universal form~(\ref{ktuniv}) is well satisfied in the weak-coupling limit ($\tilde g(T)\simeq \tilde g\lesssim 1$). In this limit, we find 
\begin{equation}
\left( \frac{\mu}{T}\right)_{\rm BKT}  \simeq \frac{0.982}{2\pi}\tilde g\ln\left(\frac{2\times 9.48}{\tilde g}\right) ,
\end{equation}
in good agreement with the weak-coupling result~\cite{Popov_book_2,Fisher88,Prokofev01,Prokofev02} 
\begin{equation}
\left( \frac{\mu}{T}\right)_{\rm BKT}  = \frac{1}{2\pi}\tilde g\ln\left(\frac{2\zeta}{\tilde g}\right), 
\label{Tktmc}
\end{equation}
where $\zeta\simeq 13.2\pm 0.4$ has been obtained from Monte Carlo simulation~\cite{Prokofev01,Prokofev02}. We ascribe the violation of universality at strong coupling, as seen in Fig.~\ref{fig_Tkt}, to a poor description of the BKT transition by the NPRG when $\tilde g\gtrsim 1$~\cite{note14}. Although we can use the same method to determine the BKT transition temperature in the Bose-Hubbard model, we cannot compare with the experimental result of the Chicago II experiment~\cite{Zhang12} which corresponds to a strong-interaction regime ($\tilde g_{\rm BH}(T)\sim 4.3$) where this method is not reliable.  

\section{Conclusion}

The scale invariance of the equation of state of a weakly interacting Bose gas, {\it i.e.} the fact that the phase-space pressure $\calP(\mu,T)$ depends only on $\mu/T$ when the dimensionless interaction constant $\tilde g$ is small, is well understood both experimentally and theoretically. We have shown that, more generally, the phase-space pressure $\calP(\mu,T)$ is a universal function of $\mu/T$ and the temperature-dependent dimensionless interaction constant $\tilde g(T)$ [Eq.~(\ref{pressure0})]. Using the NPRG approach, we have computed the corresponding universal scaling function $\calF(x,y)$ for a two-dimensional gas from weak to strong coupling. Recent measurements of the pressure, density and entropy in a weakly two-dimensional Bose gas~\cite{Yefsah11,Hung11} allow us to determine both the $x$ and $y$ dependence of $\calF(x,y)$ in some limits, and the results are found to agree remarkably well with the NPRG predictions. 

We have also compared our theoretical results in the Bose-Hubbard model with recent experimental data obtained in a two-dimensional Bose gas in an optical lattice near the vacuum-superfluid transition~\cite{Zhang12}. Our theoretical analysis shows that the lowest temperature ($T=2.5t$) reached in the experiment remains slightly above the crossover temperature $T\sim t$ to the quantum critical regime where the thermodynamics is fully determined by the universal scaling function $\calF$. However, somewhat surprisingly, the experimental data do not show the small deviations from (universal) quantum critical behavior that are expected for $T=2.5t$ (see the discussion in Sec.~\ref{sec_exp}).

The experiment reported in Ref.~\cite{Zhang12} shows that it is now possible to measure the thermodynamics of a two-dimensional Bose gas in an optical lattice near the superfluid--Mott-insulator transition (where the Mott insulating phase is not the vacuum). Since this transition (when it is induced by a density change) belongs to the dilute Bose gas universality class, the thermodynamics in the superfluid phase is also determined by the scaling function $\calF$ (the BKT transition temperature being determined by the scaling function $\calH$, see Eq.~(\ref{ktuniv})). The nonuniversal parameters  $m$ and $a_2$ should be understood as the effective mass and effective scattering length of the elementary excitations at the (nontrivial) QCP between the superfluid phase and the Mott insulator. We have recently shown that Eq.~(\ref{pressure0}) indeed holds for a three-dimensional Bose gas in an optical lattice near the Mott transition and computed the non-universal parameters $m$ and $a_3$ in the framework of the 
Bose-Hubbard model~\cite{Rancon12a}. Measuring the thermodynamics near the superfluid--Mott-insulator transition of a two- or three-dimensional Bose gas would allow for a very strong test of universality in strongly interacting quantum fluids.

\begin{acknowledgments}
We would like to thank X. Zhang, C. Chin, R. Desbuquois and J. Dalibard for discussions and/or correspondence and providing us with the experimental data shown in the manuscript. We are especially grateful to X. Zhang and C. Chin for numerous correspondences about the experiment reported in Ref.~\cite{Zhang12}. 
\end{acknowledgments} 

\appendix

\section{Non-perturbative RG}
\label{app_nprg}

In this section, we briefly review the NPRG approach to interacting boson systems. The strategy of the NPRG is to build a family of models indexed by a momentum scale $k$ varying from a microscopic scale $\Lambda$ down to 0. In practice, this is achieved by adding to the (Euclidean) action $S$ of the model a term 
\begin{equation}
\Delta S_k[\psi^*,\psi] = \inttau \sum_\q \psi^*(\q) R_k(\q) \psi(\q) .
\end{equation}
Here $\psi_\q(\tau)$ is a bosonic field and $\tau\in[0,1/T]$ an imaginary time. The so-called cutoff function $R_k(\q)$ vanishes for $k=0$ so that the action $S+\Delta S_{k=0}$ reduces to the action of the model we are interested in. When $k$ is finite, $R_k(\q)$ suppresses fluctuations with momenta $|\q|\lesssim k$ and acts as an infrared regulator term. Its value at the microscopic scale $\Lambda$ must be chosen such that the model with action $S+\Delta S_\Lambda$ is exactly (at least numerically) solvable. In the standard implementation of the NPRG~\cite{Berges02,Delamotte07}, one ensures that all fluctuations are frozen by the $\Delta S_\Lambda$ term, so that the mean-field approximation becomes exact. For a lattice model, such as the Bose-Hubbard model, one can instead choose $R_\Lambda(\q)$ (with $\Lambda$ of the order of the inverse lattice spacing) such that $t_\q+R_\Lambda(\q)=0$. The action $S+\Delta S_\Lambda$ then describes a system of decoupled sites (vanishing hopping amplitude) and is exactly 
solvable. This implementation of the NPRG, referred to as the lattice NPRG, was introduced in Ref.~\cite{Machado10} and used to study the Bose-Hubbard model in Refs.~\cite{Rancon11a,Rancon11b}. In both the standard and lattice NPRG schemes, the effective action of the original system (with action $S$) is deduced from that of the reference system (with action $S+\Delta S_\Lambda$) by solving a RG equation. 

The main quantity of interest in the NPRG approach is the scale-dependent effective action 
\begin{align}
\Gamma_k[\phi^*,\phi] ={}& - \ln Z_k[J^*,J] + \inttau \intr (J^* \phi + \cc) \nonumber \\ 
& - \Delta S_k[\phi^*,\phi] , 
\end{align}
defined as a modified Legendre transform of $-\ln Z_k[J^*,J]$ which includes the subtraction of $\Delta S_k[\phi^*,\phi]$. Here $J(\r,\tau)$ is an external (complex) source which couples linearly to the boson field $\psi(\r,\tau)$, $Z_k[J^*,J]$ the partition function corresponding to the action $S+\Delta S_k$, and 
\begin{equation}
\phi(\r,\tau) = \frac{\delta\ln Z_k[J^*,J]}{\delta J^*(\r,\tau)} = \mean{\psi(\r,\tau)}
\end{equation}
the superfluid order parameter. 

In the standard NPRG implementation, the initial value $\Gamma_\Lamb[\phi^*,\phi]=S[\phi^*,\phi]$ of the scale-dependent effective action is given by the microscopic action. In the lattice implementation (Bose-Hubbard model),
\begin{equation}
\Gamma_\Lambda[\phi^*,\phi] = \Gamma_{\rm loc}[\phi^*,\phi] + \inttau \sum_\q \phi^*(\q) t_\q \phi(\q), 
\end{equation}
where $\Gamma_{\rm loc}$ is the Legendre transform of the thermodynamic potential $-\ln Z_{\rm loc}[J^*,J]$ in the local limit (vanishing hopping amplitude). The effective action $\Gamma_k$ can be deduced from $\Gamma_\Lambda$ by (approximately) solving the exact flow equation~\cite{Wetterich93} 
\begin{equation}
\dk \Gamma_k[\phi^*,\phi] = \half \Tr\llbrace \dk R_k \left( \Gamma_k^{(2)}[\phi^*,\phi]+R_k \right)^{-1} \rrbrace ,
\label{eqwet}  
\end{equation}
where $\Gamma_k^{(2)}$ is the second-order functional derivative of $\Gamma_k$. In Fourier space, the trace in~(\ref{eqwet}) involves a sum over momenta and frequencies as well as the two components of the complex field $\phi$. We refer to Refs.~\cite{Dupuis07,*Dupuis09a,*Dupuis09b,Rancon11b,Rancon12a} for a detailed discussion of the approximations used to solve Eq.~(\ref{eqwet}). 

All thermodynamics  properties can be obtained from the effective potential defined by 
\begin{equation}
V_k(n) = \frac{1}{\beta V} \Gamma_k[\phi^*,\phi] \biggl|_{\phi\,\const}  
\end{equation}
($V$ denotes the volume of the system), where $\phi$ is a constant (uniform and time-independent) field and $n=|\phi|^2$. Its minimum determines the condensate density $n_{0,k}$ and the thermodynamic potential (per unit volume) $V_{0,k}=V_k(n_{0,k})$ in the equilibrium state. The pressure is then simply given by 
\begin{equation}
P(\mu,T) = - V_{0,k=0} . 
\end{equation}

\section{NPRG equations for $\mu=0$}
\label{app_muzero}

In this section, we discuss the solution of the $\mu=0$ NPRG equation for a continuum model with Hamiltonian~(\ref{ham}) ($d=2$). We use the cutoff function $R_k(\q)=(\eps_k-\eps_\q)\Theta(\eps_k-\eps_\q)$ ($\Theta(x)$ denotes the step function) and approximate the effective potential by 
\begin{equation}
V_k(n) = V_{0,k} + \delta_k n + \frac{\lamb_k}{2} n^2 . 
\label{trunc}
\end{equation}
The initial condition given by the mean-field solution is $\delta_\Lambda=-\mu$ and $\lamb_\Lambda=g$. The RG equations read
\begin{align}
\dk \delta_k ={}& \lamb_k \tdk \int_\q \coth\left(\frac{\eps_\q+\delta_k+R_k(\q)}{2T}\right) , \nonumber \\
\dk \lamb_k ={}& -\frac{\lamb_k^2}{2} \tdk \int_\q \biggl[\frac{1}{\eps_\q+\delta_k+R_k(\q)} \nonumber \\ & \times \coth\left(\frac{\eps_\q+\delta_k+R_k(\q)}{2T}\right) \nonumber \\
& + \frac{2}{T} \sinh^{-2} \left(\frac{\eps_\q+\delta_k+R_k(\q)}{2T}\right) \biggr] ,
\label{rgeq3} 
\end{align}
where $\tdk=(\dk R_k)\partial/\partial_{R_k}$ and we use the notation $\int_\q\equiv (2\pi)^{-2}\int d^2q$. Note that with the truncation~(\ref{trunc}), the NPRG equations in the normal phase (vanishing condensate density) reduce to standard one-loop equations. Except for minor differences (due to a slightly different RG scheme), Eqs.~(\ref{rgeq3}) are equivalent to the one-loop RG equations derived in Refs.~\cite{Fisher88,Sachdev94} using a momentum-shell one-loop RG approach with a sharp cutoff. 

\begin{figure}
\centerline{\includegraphics[width=7cm]{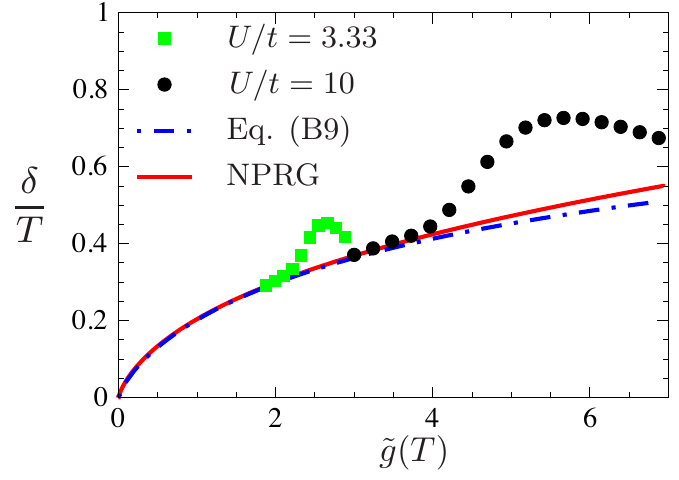}}
\caption{(Color online) $\delta/T$ vs $\tilde g(T)$ obtained from the numerical solution of Eqs.~(\ref{rgeq3}) with $\mu=0$ ((red) solid line). The (blue) dash-dotted line shows the approximate solution~(\ref{rgeq8a}) with an overall multiplicative constant $\alpha=0.87$. The symbols show $\delta/T$ in the Bose-Hubbard model for $U/t=3.33$ and $10$ (and $\mu=\mu_c$).}
\label{fig_delta}
\end{figure}

Since $\eps_\q+R_k(\q)\geq \eps_k$, the RG equations can be approximated by their $T=0$ limit when $\eps_k\gg T$, 
\begin{equation}
\begin{split}
\dk \delta_k &= 0 , \\ 
\dk \lamb_k &= \frac{m\lamb_k^2}{2\pi} .
\end{split} 
\label{rgeq4} 
\end{equation}
There is a quantum-classical crossover when $k$ becomes of the order of $k_T=\sqrt{2mT}$. In the classical regime where $k\ll k_T$, $\eps_\q+\delta_k+R_k(\q)\ll T$ for the values $|\q|\sim k$ contributing to the momentum integrals in~(\ref{rgeq3}), we find 
\begin{equation}
\begin{split}
\dk \delta_k &= - \frac{2mT}{\pi} \lamb_k , \\ 
\dk \lamb_k &= \frac{20m^2T}{\pi k^2} \lamb_k^2 . 
\end{split} 
\label{rgeq5} 
\end{equation}
To obtain an approximate solution of the RG equations, we first integrate~(\ref{rgeq4}) between $k=\Lambda$ and $k=k_T$, which gives
\begin{equation}
\begin{split}
\delta_{k_T} &\simeq 0 , \\ 
\lamb_{k_T} & \simeq g(T) ,
\end{split} 
\label{rgeq6}
\end{equation}
where $g(T)=\tilde g(T)/2m$ is defined by~(\ref{gT}). We then integrate Eqs.~(\ref{rgeq5}) between $k_T$ and $k=\sqrt{2m\delta_k}$ (the RG flow stops beyond this point) with boundary values at $k_T$ given by~(\ref{rgeq6}). We deduce the approximate expressions of $\delta\equiv\delta_{k=0}$ and $\lamb\equiv\lamb_{k=0}$, 
\begin{equation}
\begin{split} 
\delta &= - \frac{mT}{\pi B} \ln \left( \frac{g(T)}{2mT}(A+2mB\delta) \right), \\
\lamb &= \frac{2m\delta}{A+2m\delta B} , 
\end{split}
\end{equation}
where
\begin{equation}
A = \frac{10m^2T}{\pi}, \qquad  
B = \frac{1}{g(T)} - \frac{A}{2mT} . 
\label{rgeq7} 
\end{equation}
The one-loop RG equations are essentially exact in the quantum regime $k\gg k_T$ (which coincides with the vacuum limit when $\mu=0$) but requires $\tilde\lamb_k=2m\lamb_k\leq\tilde\lamb_{k_T}\ll 1$, {\it i.e.} $\tilde g(T)\ll 1$, to be valid in the classical regime. Using $\delta\ll T$ and anticipating that $\delta/T\gg \tilde g(T)$, the equation for $\delta$ simplifies into
\begin{equation}
\frac{\delta}{T} = - \frac{\tilde g(T)}{2\pi} \ln \left(\frac{\delta}{T} \right) . 
\label{rgeq8}
\end{equation}
The solution of Eq.~(\ref{rgeq8}),
\begin{equation}
\frac{\delta}{T} = \frac{\tilde g(T)}{2\pi} W\left(\frac{2\pi}{\tilde g(T)}\right) ,  
\label{rgeq8a}
\end{equation}
can be written in terms of the Lambert function $W(x)$ defined by $We^W=x$. Using $W(x)\simeq \ln x$ for $x\gg 1$, we obtain 
\begin{equation}
\frac{\delta}{T} = \frac{\tilde g(T)}{2\pi} \ln\left(\frac{2\pi}{\tilde g(T)}\right)  
\label{rgeq8b}
\end{equation}
for $\tilde g(T)\to 0$. The gap $\delta\equiv\delta_{k=0}$ obtained from the NPRG equations is shown in Fig.~\ref{fig_delta}. There is a very good agreement with the expression~(\ref{rgeq8a}) up to a multiplicative constant $\alpha\simeq 0.87$ (which accounts for the rather crude treatment of the quantum-classical crossover when solving the RG equations). Figure~\ref{fig_delta} also shows the gap computed in the Bose-Hubbard model. At low temperatures, we recover the universal limit described by the continuum model. We also observe a non-monotonous variation of $\delta/T$ which is due to the enhanced density of states of the square lattice near the band center~\cite{note11}. The maximum of $\delta/T$ with respect to $\tilde g(T)$ is responsible for the maximum observed in the phase-space pressure $\calP(\mu_c,T)$ as a function of $\tilde g(T)$ (see Fig.~\ref{fig_P_lat}). 

To obtain the pressure, we adopt the standard momentum-shell RG point of view where $-\delta$ is interpreted as a renormalized chemical potential. Computing the pressure directly from the renormalized parameters~\cite{Sachdev94}, and ignoring the renormalized interaction $\tilde\lamb \ll 1$, we obtain 
\begin{equation}
\calP(0,T) = \mbox{Li}_2(e^{-\delta/T}) , 
\label{rgeq9} 
\end{equation}
where Li$_2$ is a polylogarithm function. Together with the condition $\delta\ll T$ and the asymptotic behavior of Li$_2(x)$ for $|x-1|\ll 1$ and $x<1$, Eq.~(\ref{rgeq9}) gives 
\begin{equation}
\calP(0,T) = \frac{\pi^2}{6} -  \frac{\tilde g(T)}{2\pi} \ln^2\left(\frac{2\pi}{\tilde g(T)}\right) . 
\label{rgeq10} 
\end{equation}
This expression agrees with the numerical solution of Eqs.~(\ref{rgeq3}) but differs from the analytical result reported in Ref.~\cite{Sachdev94}. The result $\lim_{T\to 0}\calP(0,T)=\pi^2/6$ is exact since the one-loop approximation becomes exact in the limit $\tilde g(T)\to 0$. 

To compute the density $D$ and the compressibility $\kappa$, we start the RG procedure with an infinitesimal chemical potential, {\it i.e.} $\delta_\Lambda=-\mu$. Integrating the RG equations in the classical and quantum regimes, we then find $\delta_{k_T}\simeq -\mu$ and $\delta(\mu)=\delta(0)+\delta_{k_T}$, where $-\delta(0)$ is the renormalized chemical potential for $\mu=0$ [Eq.~(\ref{rgeq8a})]. We deduce 
\begin{equation}
\begin{split}
D(0,T) &= \frac{\partial P(\mu,T)}{\partial \mu}\biggl|_{\mu=0} = - \frac{\partial P(0,T)}{\partial \delta} , \\
\kappa(0,T) &= \frac{\partial^2 P(\mu,T)}{\partial \mu^2}\biggl|_{\mu=0} = \frac{\partial^2 P(0,T)}{\partial\delta^2} .
\end{split}
\end{equation}
Using $\mbox{Li}_2'(x)=-\ln(1-x)/x$, we finally obtain 
\begin{equation}
\begin{split}
\calD(0,T) &= - \ln \bigl(1-e^{-\delta/T}\bigr) \simeq \ln\left(\frac{2\pi}{\tilde g(T)}\right) , \\
\kappa(0,T) &= \frac{m}{2\pi} \frac{1}{e^{\delta/T}-1} \simeq \frac{m}{\tilde g(T)  \ln\left(\frac{2\pi}{\tilde g(T)}\right)} , 
\end{split}
\label{rgeq11} 
\end{equation}
a result which is also in good agreement with the numerical solution of the RG equations. Note that the compressibility $\kappa(0,T)$ diverges at the quantum critical point $\mu=T=0$, in agreement with the known result 
\begin{equation}
\kappa(\mu,T=0) \simeq -\frac{m}{2\pi} \ln \left(\half \sqrt{ma_2^2\mu}\right) \simeq \frac{2m}{\tilde g(\mu)} 
\end{equation}
for small positive $\mu$ (see Eq.~(\ref{P2})).

\section{Interaction constant $\gbh(\eps)$ in the Bose-Hubbard model}
\label{app_disp} 

In the continuum model, the energy-dependent interaction constant is defined from the solution of the $T=\mu=0$ RG equation. The latter can be written as
\begin{equation}
g(\eps) = \frac{g}{1+g\Pi(\eps)} , 
\label{g1}
\end{equation}
where 
\begin{align}
\Pi(\eps) &= \int_\q \intw \Theta(\eps_\q-\eps) \frac{1}{(i\w-\eps_\q)(-i\w-\eps_\q)} \nonumber \\ 
&= \int_\q \Theta(\eps_\q-\eps) \frac{1}{2\eps_\q} 
\label{g2}
\end{align}
is a particle-particle propagator with an infrared cutoff $\eps$. Equation~(\ref{g1}) makes it clear that the one-loop RG equation resums the ladder diagrams contributing to the two-particle vertex. For bosons with a quadratic dispersion $\eps_\q=\q^2/2m$ (and an energy-independent density of states $N(\eps)=\int_\q \delta(\eps-\eps_\q)=m/2\pi$), Eq.~(\ref{g2}) is equivalent to 
\begin{equation}
\Pi(\eps) = \int_\q \frac{1}{2(\eps_\q+\eps)} 
\label{g3}
\end{equation}
in the limit $\eps\ll\Lambda^2/2m$. In the Bose-Hubbard model, we define the energy-dependent interaction constant $\gbh(\eps)$ from~(\ref{g1}) and (\ref{g3}) with $g\equiv U$ and $\eps_\q$ the lattice dispersion~(\ref{epsq}) [see Eqs.~(\ref{gbh},\ref{pibh})]. Note that using~(\ref{g2}) rather than (\ref{g3}) would yield the same universal limit~(\ref{gT}) in the low-energy limit.


\begin{thebibliography}{68}
\bibitem [{not({\natexlab{a}})}]{note2}%
  \BibitemOpen
  \href@noop {} {} {\natexlab{a}} \bibinfo {note} {Although irrelevant in
  the renormalization-group sense, the boson-boson interaction constant
  necessary enters the equation of state since an ideal Bose gas is unstable
  for $\mu>0$. Furthermore, $\tilde g(T)$ is a dougerously irrelevant variable
  above the upper critical dimension $d_c^+=2$, which means that the scaling
  function $\calF_d$ [Eq.~(\ref{pressure0})] may depend on $\tilde g(T)$ in a
  singular way.}\BibitemShut {Stop}%
\bibitem [{not({\natexlab{b}})}]{note15}%
  \BibitemOpen
  \href@noop {} {} {\natexlab{b}} \bibinfo {note} {In a three-dimensional
  Bose gas, $l$ is given by the van der Waals length $\lvdw$, which is
  typically of the order of the $s$-wave scattering length $a_3$. In a
  quasi-two-dimensional gas, it is given by the thickness $l_z$ of the gas in
  the confining direction (see Sec.~\ref{sec_model}).}\BibitemShut {Stop}%
\bibitem [{not({\natexlab{d}})}]{note1}%
  \BibitemOpen
  \href@noop {} {} {\natexlab{d}} \bibinfo {note} {Since the QCP ground
  state is the vacuum, a particle added to the ground state propagates as a
  free particle. The single-particle propagagor is therefore given exactly by
  the non-interacting propagator [Eq.~(\ref{propa})].}\BibitemShut {Stop}%
\bibitem [{not({\natexlab{f}})}]{note3}%
  \BibitemOpen
  \href@noop {} {} {\natexlab{f}} \bibinfo {note} {The dimensionless
  variables $\tilde T$, $\tilde\mu$ and $\tilde P$ are defined according to
  their scaling dimensions at the fixed point: $[\tilde T]=z$, $[\tmu]=1/\nu$
  and $[\tilde P]=d+z$. With the additional $2m$ factors (and the fact that
  $z=1/\nu=2$ at the Gaussian fixed point), these quantities have no physical
  units.}\BibitemShut {Stop}%
\bibitem [{not({\natexlab{h}})}]{note8}%
  \BibitemOpen
  \href@noop {} {} {\natexlab{h}} \bibinfo {note} {When these higher-order
  contributions are taken into account, the scaling function $\calF$ must
  involve additional arguments, such as the dimensionless three-body
  interaction constant, and the low-energy physics cannot be described in terms
  of the mass $m$ and the scattering length $a_d$ only. In two dimensions
  however, it turns out that all $n$-body interaction constants ($n\geq 2$) can
  be expressed as a function of $a_2$ (see Ref.~\cite{Adhikari95}) and the
  universal scaling function $\calF_2$ remains a function of $\mu/T$ and
  $a_2$.}\BibitemShut {Stop}%
\bibitem [{not({\natexlab{i}})}]{note5}%
  \BibitemOpen
  \href@noop {} {} {\natexlab{i}} \bibinfo {note} {Noting that $\tilde
  g(\eps)$ is a function of $ma_d^2\eps$, the universal scaling function
  $\calG_d$ can be easily related to $\calF_d$.}\BibitemShut {Stop}%
\bibitem [{not({\natexlab{j}})}]{note10}%
  \BibitemOpen
  \href@noop {} {} {\natexlab{j}} \bibinfo {note} {The temperature for
  which $\tilde g(T)=\tilde g/2$, $T=(2/ma_2^2)\exp(-2C-16\pi/\tilde
  g)=(\Lamb^2/2m)(\Lamb a_2/2)^2 e^{2C}$, is proportionnal to
  $a_2^2$.}\BibitemShut {Stop}%
\bibitem [{not({\natexlab{k}})}]{note11}%
  \BibitemOpen
  \href@noop {} {} {\natexlab{k}} \bibinfo {note} {The density of states of
  the square lattice is given by $\calN(\eps)=\int_\q \delta(\eps-\eps_\q)=
  (2\pi^2 t)^{-1}K[\sqrt{(\eps-4t)^2}/4t]$ ($0\leq\eps\leq 8t$) [with $\eps_\q$
  defined in Eq.~(\ref{epsq})] . The complete elliptic integral of the first
  kind $K(x)$ diverges logarithmically for $x\to 0$.}\BibitemShut {Stop}%
\bibitem [{not({\natexlab{l}})}]{note17}%
  \BibitemOpen
  \href@noop {} {} {\natexlab{l}} \bibinfo {note} {Note that the divergence
  of the compressibility $\kappa$ at the quantum critical point $\mu=T=0$ was
  overlooked in Ref.~\cite{Fang11}.}\BibitemShut {Stop}%
\bibitem [{not({\natexlab{m}})}]{note6}%
  \BibitemOpen
  \href@noop {} {} {\natexlab{m}} \bibinfo {note} {Our definition of the
  dimensionless interaction constant $\tilde g=2mg$ differs by a factor of 2
  from that used in Refs.~\cite{Yefsah11,Hung11,Zhang12}.}\BibitemShut {Stop}%
\bibitem [{not({\natexlab{p}})}]{note14}%
  \BibitemOpen
  \href@noop {} {} {\natexlab{p}} \bibinfo {note} {While the beta function
  $\beta(n_{s,k})=k\dk n_{s,k}$ for $n_{s,k}>n_s^*$ is very strongly suppressed
  at low temperatures in the weak-interaction regime, when $\tilde g(T)\gtrsim
  1$ its value significantly increases and the NPRG description of the BKT
  transition is not reliable any more.}\BibitemShut {Stop}%
\end{thebibliography}

%

\end{document}